\documentclass[10pt]{iopart}
\usepackage{iopams}
\usepackage{accents}
\usepackage{dsfont}
\usepackage{mathrsfs}
\usepackage[T1]{fontenc}
\usepackage[latin1]{inputenc}
\usepackage{calrsfs}
\usepackage[perpage,flushmargin]{footmisc}
\DefineFNsymbols*{lamport}{{${}^{*}$}{${}^{**}$}{${}^{***}$}\P\|%
                   {**}{\dagger\dagger}{\ddagger\ddagger}}
\setfnsymbol{lamport}
\usepackage{pstool}
\usepackage{graphicx}

\usepackage{color}
\usepackage{hyperref}
\usepackage{cite}
\newcommand{\utilde}[1]{\undertilde{#1}}
\newcommand{\di}{\mathrm{d}}
\usepackage{tensor}
\newcommand{\ou}[3]{\tensor{#1}{^{#2}_{#3}}}
\newcommand{\uo}[3]{\tensor{#1}{_{#2}^{#3}}}
\newcommand{\vo}[3]{\tensor[^{\mathnormal{#2}}]{#1}{^{#3}}}
\newcommand{\vu}[3]{\tensor[^{\mathnormal{#2}}]{#1}{_{#3}}}
\newcommand{\vou}[4]{\tensor[^{\mathnormal{#2}}]{#1}{^{#3}_{#4}}}

\newcommand{\I}{\mathrm{i}} 
\newcommand{\E}{\mathrm{e}} 
\newcommand{\C}{\mathbb{C}}

\newcommand{\R}{\mathbb{R}}
\newcommand{\Z}{\mathbb{Z}}

\usepackage{slashed}
\newcommand{\varSigma}{\mathit{\Sigma}}
\newenvironment{subalign}{\numparts\begin{eqnarray}}{\end{eqnarray}\endnumparts}

\begin{document}
\title[Angular momentum and centre of mass from generalised Witten equations]{Quasi-local gravitational angular momentum and centre of mass from generalised Witten equations}
\author{Wolfgang Wieland}
\address{Perimeter Institute for Theoretical Physics\\31 Caroline Street North\\ Waterloo, ON N2L\,2Y5, Canada
}

\begin{abstract}
Witten's proof for the positivity of the ADM mass gives a definition of energy in terms of three-surface spinors. In this paper, we give a generalisation for the remaining six Poincaré charges at spacelike infinity, which are the angular momentum and centre of mass. The construction improves on certain three-surface spinor equations introduced by Shaw. We solve these equations asymptotically obtaining the ten Poincaré charges as integrals over the Nester\,--\,Witten two-form. We point out that the defining differential equations can be extended to three-surfaces of arbitrary signature and we study them on the entire boundary of a compact four-dimensional region of spacetime. The resulting quasi-local expressions for energy and angular momentum are integrals over a two-dimensional cross-section of the boundary. For any two consecutive such cross-sections, conservation laws are derived that determine the influx (outflow) of matter and gravitational radiation.
						
\end{abstract}
\pacs{04.20.Gz, 04.20.Ha}

\section{Introduction}\label{sec1}


Observables are defined by measurement protocols, and for local observables this must include a description for how to measure space, time, distance and duration. In field theory over Minkowski space, we can easily construct such observables from the matter fields and their derivatives at a point. 
What makes this construction physically meaningful is the presence of an absolute structure (the inertial reference frames) with respect to which we can localise fields and measure time and distance. No such absolute structure exists in full general relativity, since the outcomes for measurements of space and time depend themselves on the strength of the gravitational field, which, in turn, spoils the construction of local observables. 

\paragraph{} If spacetime is asymptotically flat, the simplest observables are the total energy, the linear momentum, the angular momentum and the centre of mass \cite{ADMmass,Regge:1974zd, Beig:1987zz} due to Arnowitt, Deser, Misner (ADM) and Beig and Ó\,Murchadha \cite{Beig:1987zz}. The \textsc{ADM} energy is always positive, as proven by Schoen and Yau \cite{Schon:1979rg}. Witten \cite{Wittenproof} developed a much simpler argument for the positivity of the \textsc{ADM} energy, which is based on spinors that solve the three-dimensional Dirac equation $\slashed{D}\psi=0$ on a Cauchy hypersurface $\varSigma$, with $\slashed{D}$ denoting the three-dimensional Dirac operator with respect to the Ashtekar\,--\,Sen connection \cite{Newvariables, Sen:1982qb}. Nester \cite{Nester:1982tr} found a fully covariant description of Witten's formalism. In Nester's approach the spatial slice $\varSigma$ could have arbitrary extrinsic curvature, while in Witten's original proof $\varSigma$ was implicitly assumed to have no extrinsic curvature. 

Witten's approach is manifestly background invariant. The solutions of the three-dimensional Dirac equation $\slashed{D}\psi=0$ are simply inserted into the asymptotic integral over the Nester\,--\,Witten two-form $e_\alpha\wedge \bar{\psi}\gamma^\alpha  D\psi$, which returns the \textsc{ADM} energy for an inertial observer at spacelike infinity, whose four-velocity is $\bar{\psi}\gamma^\mu\psi\,\uo{e}{\mu}{a}$ (the integral exists provided certain falloff conditions for the tetrad $\uo{e}{\mu}{a}$ and connection are satisfied).  The asymptotic rest frames enter Witten's construction only through the perturbative expansion of the integral $\int_{\partial\varSigma}e_\alpha\wedge \bar{\pi}\gamma^\alpha  D\pi$ at large distance, the definition itself is completely coordinate invariant.

\paragraph{} At finite distance the situation is more complicated. There are various different approaches. 
Szabados \cite{Szabados:2004vb} gives a broad review over all the current quasi-local ideas and constructions.  In Penrose's approach \cite{penroserindler2}, for example, the gravitational charges are two-dimensional curvature\footnote{The curvature two-form $F_{CDab}$ is the selfdual component of the Riemann curvature tensor $R_{cdab}$.} integrals $\int_S F_{AB}\phi^A\psi^B$, where the Weyl spinors $\phi^A$ and $\psi^A$ are solutions of a certain two-surface twistor equation intrinsically defined on $S$, which is a spacelike two-surface. Dougan and Mason \cite{MasonEnergy} gave a similar quasi-local definition of the gravitational energy (and angular momentum) in terms of holomorphic spinors (and twistors) on a spacelike two-surface $S$ (generically a two-sphere). The corresponding quasi-local charges are again two-dimensional surface integrals. 
  A detailed discussion of the approach by Dougan and Mason for both energy and angular momentum can be found in a series of articles by Szabados \cite{SzabadosPositivity, SzabadosSen}. Bramson \cite{BramsonSpin} gave a proposal for the \emph{intrinsic} gravitational spin in contrast to the \emph{orbital} angular momentum. His integral is $\int_S\Sigma_{AB}\lambda^A\lambda^B$, where $\Sigma_{AB}$ is the selfdual component of the Pleba\'{n}ski two-form $\Sigma_{\alpha\beta}=e_\alpha\wedge e_\beta$ and the spinor $\lambda^A$ is a solution of a certain asymptotic twistor equation \cite{BramsonFrame} around null infinity. This does not seem to be the only reasonable choice for $\lambda^A$. Ludvigsen and Vickers \cite{LudvigsenAnglrMntm} gave a different proposal, which is discussed in great detail in \cite{SzabadosSmall,SzabadosLarge}. Another possibility is to demand that $\lambda^A$ is holomorphic \cite{SzabadosSmall,SzabadosLarge,Szabados_pp} on $S$, which follows Dougan and Mason's construction \cite{MasonEnergy}. There are, however, always asymptotically constant spinors $\lambda^A:\ou{\sigma}{A}{Ba}h^{ab}D_b\lambda^B=0$ for which Bramson's quasi-local angular momentum will vanish, (we will see this explicitly in equation \eref{vanshbulk} below). 

\paragraph{} There is good reason to study such quasi-local charges. Spacelike infinity is a mathematical idealisations, which crucially requires the vanishing of the cosmological constant \cite{Ashtekar:2014zfa}. Any realistic measurement is taken at large but finite distance from the source. 
 How do we then define energy and angular momentum at finite distance?  The key message of this paper is that a suitable generalisation of Witten's approach offers a systematic way to construct such quasi-local observables for both energy and angular momentum. The proposal is defined through certain first-order partial differential equations for spinors on a three-surface, which may be spacelike, time-like or null. 

This three-surface is closed and splits the gravitational degrees of freedom into an external environment (\emph{the observer}) and the system observed, which is a four-dimensional spacetime region $\mathcal{B}$ with a three-dimensional boundary (see figure \ref{fig1} for an illustration). The quasi-local observables \emph{for this choice of observer} are then surface integrals over two-dimensional cross-sections $S$ of the boundary $\partial\mathcal{B}$. At finite distance from the source, there are infinitely many such obervables, but at spacelike infinity only ten of them survive, which are the ADM energy-momentum, the angular momentum, and the centre of mass (all the other charges either diverge or vanish in the limit of infinite distance from the source). What makes the construction attractive is that these infinitely many charges are subject to infinitely many conservation laws that determine the influx (outflow) of matter and gravitational radiation across the boundary.\footnote{We study the evolution of the charges along a foliation of the three-boundary. To speak about evolution in any sensible way this foliation should define a causal vector-field (i.e.\ future oriented, time-like or null) along a region of the three-boundary. In this paper, we will content ourselves with the most relevant case, which is when this vector field is null.}

\paragraph{} As a first step in this program, we need to generalise Witten's original construction. Witten's proof was tailored to the \textsc{ADM} linear momentum. For the remaining six Poincaré charges, which are the angular momentum and the centre of mass, we need to introduce a suitable generalisation of Witten's equation. 
The system of equations that we will study for this purpose has been introduced before by Shaw \cite{0264-9381-2-2-012} in 1985. In his paper, Shaw was mostly concerned with angular momentum at null infinity, and he was studying spatial hypersurfaces, which are asymptotically null. He gave a heuristic discussion of spacelike infinity as well, but a detailed analysis of the asymptotics was missing. The present paper closes this gap and shows that Shaw's generalised Witten equations return all ten Poincaré charges at spacelike infinity, provided the usual parity and falloff conditions due to Regge and Teitelboim \cite{ReggeTeitelboim} are satisfied. We will then also point out that Shaw's generalised Witten equations can be extended to three-surfaces of arbitrary signature. This is an important observation and it allows us to define a family of infinitely many quasi-local charges, which satisfy balance laws describing the influx of matter and radiation. We will consider, in particular, a four-dimensional region bounded by an expanding null surface. The quasi-local charges will be surface integrals over two-dimensional cross-sections of the boundary and we will show that the quasi-local energy can only increase between two consecutive such cross sections.

\section{Generalised Witten equation}\label{sec2}
\subsection{Elementary preparations} Witten starts his celebrated proof \cite{Wittenproof} of the positivity of the \textsc{ADM} mass from the three-dimensional Dirac equation
\begin{equation}
\ou{\sigma}{A}{Ba}h^{ab}D_b\pi^B\equiv\ou{D}{A}{B}\pi^B=0\label{Witteq0}
\end{equation}
on a Cauchy hypersurface $\varSigma$ of an asymptotically flat spacetime manifold $(M,g_{ab})$ admitting a global spin structure: $\pi^A$ is a section of the spinor bundle over the spatial hypersurface $\varSigma$ and $D_a$ denotes the pullback of the four-dimensional covariant derivative\footnote{The covariant derivative $\nabla_a$ acts on all indices, it is torsionless and annihilates both the metric $g_{ab}$ as well as $\epsilon_{AB}$, $\bar\epsilon_{A'B'}$  and the soldering form $\ou{\sigma}{AA'}{a}$.} $\nabla_a$ to $\varSigma$. Furthermore, $\ou{\sigma}{A}{Ba}$ are the Pauli matrices on $\varSigma$ and $h_{ab}=n_an_b+g_{ab}$ is the spatial three-metric, with $n^a$ denoting the future oriented hypersurface normal.

The $\mathfrak{su}(2)_n$ valued one-forms $\ou{\sigma}{A}{Ba}$ can be constructed explicitly from the four-dimensional soldering forms $\ou{\sigma}{AA'}{a}$ by setting
\begin{equation}
\ou{\sigma}{A}{Ba}:=\uo{h}{a}{b}\ou{\sigma}{AC'}{b}\delta_{BC'}=-\uo{h}{a}{b}\delta^{AC'}\bar{\sigma}_{C'Bb},\label{pmatrx}
\end{equation}
where we have introduced the Hermitian metric
\begin{equation}
\delta_{AA'}=\sigma_{AA'a}n^a.
\end{equation}
As already indicated, the one-forms $\ou{\sigma}{A}{Ba}$ on $\varSigma$ are nothing but a covariant version of Pauli's spin matrices: They are traceless, they are Hermitian with respect to the metric $\delta_{AA'}$ and they form a basis in the Lie algebra $\mathfrak{su}(2)_n$ of $SL(2,\C)$ transformations preserving $\delta_{AA'}$. They also satisfy the generalised Pauli identity
\begin{equation}
\ou{\sigma}{A}{Ca}\ou{\sigma}{C}{Bb}=\delta^A_B h_{ab}+\I\uo{\varepsilon}{ab}{c}\ou{\sigma}{A}{Bc},\label{paulident}
\end{equation}
with $\varepsilon_{abc}=n^d\varepsilon_{dabc}$ denoting the three-dimensional Levi-Civita tensor on $\varSigma$.

The covariant derivative $D_a$ defines the selfdual Ashtekar\,--\,Sen \cite{Newvariables, Sen:1982qb,AshtekarHorowitz} connection $\ou{A}{A}{Ba}$, which is the pull-back of the four-dimensional spin connection to $\varSigma$. The difference between the intrinsic three-dimensional spin connection $\ou{\Gamma}{A}{Ba}$, which is an $\mathfrak{su}(2)_n$ connection on $\varSigma$, and the Ashtekar\,--\,Sen connection  $\ou{A}{A}{Ba}$ is measured by the extrinsic curvature tensor $K_{ab}=\uo{h}{a}{c}\nabla_cn_b$ through
\begin{equation}
\ou{A}{A}{Ba}=\ou{\Gamma}{A}{Ba}+\frac{1}{2}\ou{\sigma}{A}{Bb}\ou{K}{b}{a},\label{AshSen}\\
\end{equation}
hence $\ou{A}{A}{Ba}$ is an $\mathfrak{sl}(2,\C)$ connection. Its field strength $\ou{F}{AB}{ab}=2\partial_{[a}{A}^{AB}{}_{b]}+2{A}^{A}{}_{C[a}{A}^{CB}{}_{b]}$ is given by the spatial components of the selfdual part of the four-dimensional Riemann curvature tensor $R_{abcd}$. Indeed\begin{equation}
\ou{F}{A}{Bab}=\frac{1}{2\I}\ou{\sigma}{A}{Bc}\left(\frac{1}{2}\ou{\varepsilon}{c}{de}\ou{R}{ed}{a'b'}+\I\ou{R}{c}{da'b'}n^d\right)\uo{h}{a}{a'}\uo{h}{b}{b'}.
\end{equation}

\subsection{Witten's integral} To prove the positive energy theorem, Witten inserts the solutions of the differential equation \eref{Witteq0} into the integral
\begin{equation}
\lim_{r\rightarrow\infty}\int_{S^2_r}d^2v^a\,\bar{\pi}^{A'}\delta_{AA'}D_a\pi^A\label{Wittint}
\end{equation}
over the two-sphere at spacelike infinity\footnote{The surface element is $d^2v^a=\frac{1}{2}\tilde{\eta}^{bc}\uo{\varepsilon}{bc}{a}$ with $\tilde{\eta}^{ab}=-\tilde{\eta}^{ba}$ denoting the canonical Levi-Civita surface density on the two-sphere $S^2_r$ at fixed radial coordinate $r=\sqrt{\delta_{ij}x^ix^j}$, with $\{x^i\}$ denoting an asymptotically Cartesian coordinate system on $\varSigma$.}. He then continues to show that there are solutions of the defining differential equation \eref{Witteq0} that admit the asymptotic expansion
\begin{equation}
\pi^A(\vec{x})=\vo{\pi}{0}{A}+O(r^{-1}),\label{Wittsol1}
\end{equation}
in an asymptotic Cartesian coordinate system $\{x^i\}$. The zeroth order is asymptotically constant with respect to the flat spin connection associated to $\{x^i\}$, that is $\partial_a\vo{\pi}{0}{A}=0$, and the next to leading order vanishes as $r^{-1}$ in the limit of $r=\sqrt{x^ix_j}$ going to infinity. We will repeat the argument in the next section, because we also need to show that the next to leading order is of definite parity provided the leading terms of the three-metric and Ashtekar\,--\,Sen connection satisfy certain falloff and parity conditions. The solutions \eref{Wittsol1} are then inserted into the integral. The integral converges and returns the \textsc{ADM} four-momentum at spacelike infinity
\begin{equation}\fl\quad
\lim_{r\rightarrow\infty}\int_{S^2_r}d^2v^a\,\bar{\pi}^{A'}\delta_{AA'}D_a\pi^A=4\pi G\lim_{r\rightarrow\infty}\int_{S^2_r}\tilde{P}_a\uo{\sigma}{AA'}{a}\,\vo{\pi}{0}{A}\vo{\bar\pi}{0}{A'},
\end{equation}
where $\tilde{P}_a$ is the corresponding four-momentum density.
The right hand side represents, therefore, the energy for an observer at spacelike infinity, whose four-velocity is $\ell^a=-\uo{\sigma}{AA'}{a}\pi^A\bar\pi^{A'}$. Gauss's divergence theorem turns the surface integral \eref{Wittint} into a volume integral. If one then finally inserts the field equations for a given stress-energy tensor $T_{ab}$ one arrives at
\begin{equation}\fl\quad
\lim_{r\rightarrow\infty}\int_{S^2_r}d^2v^a\,\bar{\pi}^{A'}\delta_{AA'}D_a\pi^A=\int_{\varSigma} d^3v\left[h^{ab}\delta_{AA'}D_a\pi^AD_b\bar\pi^{A'}+4\pi G T_{ab}n^a\ell^b\right]\label{postvty},
\end{equation}
with $d^3v=\frac{1}{3!}\tilde{\eta}^{abc}\varepsilon_{abc}$ denoting the canonical three-volume element on $\varSigma$. The first term is always positive and so is the second term provided the dominant energy condition is satisfied. Moreover, the whole expression vanishes if and only if $g_{ab}$ is the Minkowski metric, which are the two main results of the celebrated positive energy theorem\,---\,the \textsc{ADM} mass is always positive and vanishes only in Minkowski space.
\subsection{Generalised Witten equations} 
We now wish to generalise the original argument to express the remaining six Poincaré charges\,---\,the angular momentum and centre of mass of a gravitating system\,---\,in terms of Witten spinors. To this end, we will consider the following system of equations
\begin{subalign}
\ou{\sigma}{A}{Ba}h^{ab}D_b\pi^B=0,\label{witteq1}\\
\ou{\sigma}{A}{Ba}h^{ab}D_b\omega^B+3\ou{\delta}{A}{A'}\bar{\pi}^{A'}=0.\label{witteq2}
\end{subalign}\noindent
Compared to Witten's original construction, the only novelty concerns the second equation \eref{witteq2}, which is the three-dimensional Dirac equation sourced by the solution of the first equation. This system of elliptic partial differential equations has been studied first by Shaw \cite{0264-9381-2-2-012}, who was mainly interested in a definition of angular momentum and centre of mass at null infinity. The present paper improves on that and provides a generalisation. First of all, we will see that the construction extends to all ten Poincaré charges at spacelike infinity. Shaw gave some heuristic arguments, why this should be true, but he did not give an explicit asymptotic analysis. We will provide such an analysis in \hyperref[sec3]{section 3}, and we will then also see that the system of equations \eref{witteq1} and \eref{witteq2} can be naturally extended to three-surfaces of arbitrary signature. This will allow us to to study balance laws for a family of infinitely many quasi-local charges.

Given a solution to \eref{witteq1} and \eref{witteq2}, we will study the integral of the Nester\,--\,Witten two-form over the boundary of $\varSigma$, as defined by
\begin{equation}
Q_{\pi,\omega}:=\lim_{r\rightarrow\infty}\int_{S^2_r}d^2v^a\,\bar{\omega}^{A'}\delta_{AA'}D_a\pi^A,\label{Qcharge}
\end{equation}
and we will show that it approaches a linear combination of all ten Poincaré charges at spacelike infinity. 
To get an intuition for the generalised Witten equations \eref{witteq1} and \eref{witteq2}, let us first study them in flat space.
\subsection{Flat space solutions} Consider thus Minkwoski space $(\R^4,\eta_{ab})$ with inertial coordinates $\{x^\mu\}=\{x^0,x^i\}$ and let $\varSigma$ be the three-dimensional $x^0=0$ hypersurface. Go to momentum space, and consider the momentum eigenfunctions $\pi^A(\vec{x})=\vo{\pi}{0}{A}\,\E^{\I\vec{k}\cdot\vec{x}}$ for some fixed constant spinor $\vo{\pi}{0}{A}\in\C^2$. The matrix $\vec{k}\cdot\vec{\sigma}$ has eigenvalues $\pm|\vec{k}|$, hence the only regular solutions of the first spinor equation \eref{witteq1} in flat space are spinors $\pi^A(\vec{x})=\vo{\pi}{0}{A}$ constant in space. We  insert them into the second equation \eref{witteq2}, integrate them and find the general solution $\omega^A(\vec{x})=\vo{\omega}{0}{A}+\ou{\sigma}{AA'}{i}x^i\,\vu{\bar\pi}{0}{A'}$ for some integration constant $\vo{\omega}{0}{A}\in\C^2$. This is nothing but Penrose's twistor equation\,---\,any regular solution of the generalised Witten equations (\ref{witteq1}, \ref{witteq2}) in flat space defines a twistor $Z(\vec{x})=(\bar{\pi}_{A'}(\vec{x}),\omega^A(\vec{x}))$, which is incident to a point $\vec{x}$ on the $x^0=0$ hypersurface $\varSigma$.
 
\section{Perturbative expansion for large distance}\label{sec3}
\subsection{Perturbative expansion of the triad and connection}\label{sec31}  In solving the differential equations \eref{witteq1} and \eref{witteq2} to leading order in $r$, we first need to specify the falloff conditions of the triad and Ashtekar\,--\,Sen connection around spacelike infinity. We introduce asymptotically Cartesian coordinates $\{x^i\}$ on the initial hypersurface $\varSigma$, and expand the triad $\ou{e}{i}{a}$ around the fiducial one-forms $\di x^i_a$. We thus write
\begin{equation}
\ou{e}{i}{a}=\di x^i_a+\ou{f}{i}{a}+o(r^{-1}),\label{taylortriad}
\end{equation}
where $o(r^{-n})$ means $\lim_{r\rightarrow\infty}r^no(r^{-n})=0$, whereas $O(r^{-n})$ denotes terms that fall off like $r^{-n}$ or faster. The next to leading order term $\ou{f}{i}{a}$ is required to be $O(r^{-1})$ and parity even with respect to the asymptotic spherical coordinates $(x^1,x^2,x^3)=r(\sin\vartheta\cos\varphi,\dots)$, we have
\begin{equation}
\ou{f}{i}{a}(r,\vartheta,\varphi)=\ou{f}{i}{a}(r,\pi-\vartheta,\varphi+\pi).\label{fpari}
\end{equation}
The parity condition is required to remove divergencies and logarithmic ambiguities from the definition of the \textsc{ADM} charges \cite{ReggeTeitelboim}. Having defined a background structure, we can now use the flat Euclidean metric $\delta_{ij}$ and the corresponding fiducial triads $\di x^i_a$ and $\partial_i^a$ to raise and lower the corresponding indices. We can thus define the tensor $f_{ij}=\delta_{im}\ou{f}{m}{b}\partial^b_j$, which we can always assume to be symmetric, hence
\begin{equation}
f_{ij}=f_{ji}=O(r^{-1}).
\end{equation}
We split off the leading $r$ dependence, and write
\begin{equation}
f_{ij}(\vec{x})=\frac{\vu{f}{1\!}{ij}(\vartheta,\varphi)}{r}+o(r^{-1}).\label{Taylortriad}
\end{equation}

We then also need the expansion of the three-dimensional spin connection $\ou{\Gamma}{i}{a}$. We find it from the the torsionless condition $T^{i}{}_{ab}=2\partial_{[a}e^i{}_{b]}+2\ou{\epsilon}{i}{lm}\Gamma^{l}{}_{[a}e^{i}{}_{b]}$, where the difference tensor $\ou{\Gamma}{i}{a}$ measures the deviation of the intrinsic three-dimensional covariant derivative from $\partial_a$, which is the metric compatible torsionless derivative with respect to the flat background metric $\delta_{ab}=\delta_{ij}\di x^i_a\di x^j_b$. The expansion of the difference tensor $\ou{\Gamma}{i}{j}=\ou{\Gamma}{i}{a}\partial^a_j$ gives
\begin{subalign}
\vou{\Gamma}{\phantom{2}}{i}{j}=\frac{\vou{\Gamma}{2}{i}{j}(\vartheta,\varphi)}{r^2}+o(r^{-2}),\qquad r\longrightarrow\infty,\\
\vou{\Gamma}{2}{i}{j}=\lim_{r\rightarrow\infty}\left(r^2\epsilon^{imn}\partial_mf_{jn}\right).
\end{subalign}\noindent
where the leading term is assumed of order $O(r^{-2})$ and parity odd, $\vou{\Gamma}{2}{i}{j}(\vartheta,\varphi)=-\vou{\Gamma}{2}{i}{j}(\pi-\vartheta,\varphi+\pi)$. 
Finally, we also need the expansion for the Ashtekar\,--\,Sen connection \eref{AshSen}, which is a combination of the intrinsic spin connection $\ou{\Gamma}{i}{a}$ and the extrinsic curvature tensor $\ou{K}{i}{a}=\ou{e}{i}{b}{h}_{a}{}^{c}\nabla_c n^b$. For the \textsc{ADM} angular momentum to be well-defined, and to remove logarithmical divergencies
, we have to assume that the connection admits the asymptotic expansion
\begin{equation}
\ou{A}{A}{Ba}=
\frac{1}{2\I}\ou{\sigma}{A}{Bi}\underbrace{\left[\frac{\vou{A}{2\!}{i}{j}(\vartheta,\varphi)}{r^2}+o(r^{-2})\right]}_{\ou{A}{i}{j}=\ou{\Gamma}{i}{j}+\I\ou{K}{i}{j}}\di x^j_a.\label{Taylorashtekar}
\end{equation}
where $\ou{\sigma}{A}{Bi}$ are the usual Pauli matrices: $\sigma_i\sigma_j=\delta_{ij}\mathds{1}+\I\uo{\epsilon}{ij}{k}\sigma_k$. The leading $O(r^{-2})$ term $\vou{A}{2\!}{i}{j}(\vartheta,\varphi)$ is required to be parity odd. To summarise, the parity conditions are the following: The leading $O(r^{-1})$ order of the metric perturbation $\vou{f}{1}{i}{j}$ is even, while the leading $O(r^{-2})$ term $\vou{A}{2}{i}{j}$ in the perturbative expansion of the Ashtekar\,--\,Sen connection is odd.

\subsection{Perturbative solution of the first spinor equation}\label{sec3.2}
At the end of section \ref{sec2}, we found the general solution of the spinor equations (\ref{witteq1}, \ref{witteq2}) in flat space. We now want to expand around these flat space solutions and solve the equations perturbatively in $r^{-1}$. Let us start with the first equation \eref{witteq1}. Consider the ansatz
\begin{equation}
\pi^A(\vec{x})=\vo{\pi}{0}{A}+\frac{\vo{\pi}{1}{A}(\vartheta,\varphi)}{r}+o(r^{-1}),\qquad r\longrightarrow\infty,\label{taylorpi}
\end{equation}
where the leading order $\vo{\pi}{0}{A}$ is a spinor, which is constant with respect to the fiducial 
flat covariant derivative $\partial_a$ at spacelike infinity and the next to leading term is of order $r^{-1}$. We also require that the first two terms solve the differential equation to leading order, hence we want to find a spinor $\vo{\pi}{1}{A}(\vartheta,\varphi)$ on the sphere such that
\begin{equation}
\ou{\sigma}{A}{Ba}h^{ab}D_a\left[\vo{\pi}{0}{B}+\frac{\vo{\pi}{1}{B}(\vartheta,\varphi)}{r}\right]=o(r^{-2}).\label{orderpi1}
\end{equation}
Solving this equation to leading order in $r$, we drop all terms that vanish faster than $r^{-2}$. This brings us to the following differential equation on the sphere,
\begin{equation}
\left(-\ou{\sigma}{A}{Bi}\partial^i_r+\ou{\hat{\partial}}{A}{B}\right)\vo{\pi}{1}{B}(\vartheta,\varphi)=
-\frac{1}{2\I}\ou{\big[\sigma^i\sigma_j\big]}{A}{B}\vou{A}{2\!}{j}{i}(\vartheta,\varphi)\vo{\pi}{0}{B},\label{pieq}
\end{equation}
where we have defined the transversal components
\begin{equation}
\ou{\hat{\partial}}{A}{B}:=r\ou{\sigma}{A}{Bi}q^{ij}\partial_j,\label{transdirac}
\end{equation}
of the three-dimensional Dirac operator. Furthermore, $q^{ij}$ is the two-dimensional metric
\begin{equation}
q^{ij}=\delta^{ij}-\partial^i_r\partial^j_r,
\end{equation}
with $\partial^i_r=x^i/r$ denoting the outwardly oriented unit normal (with respect to the flat fiducial thee-metric).
The transversal Dirac operator $\ou{\hat{\partial}}{A}{B}$ can be immediately diagonalised in terms of the orbital $L_i=-\I\uo{\epsilon}{ij}{k}x^j\partial_k$ and spin angular momentum $S_i=\frac{1}{2}\sigma_i$ operators. We multiply $\ou{\hat{\partial}}{A}{B}$ by $\ou{\sigma}{A}{Bi}\partial^i_r$ from the left and use the Pauli identity $\sigma_i\sigma_j=\delta_{ij}\mathds{1}+\I\uo{\epsilon}{ij}{k}\sigma_k$ to get
\begin{eqnarray}\nonumber\fl\qquad
\partial^i_r\ou{\sigma}{A}{Ci}\ou{\hat{\partial}}{C}{B}&=r\partial^i_r\ou{\big[\sigma_i\sigma_j\big]}{A}{B}q^{jk}\partial_k=\I\uo{\epsilon}{i}{jk}\ou{\sigma}{A}{Bk}x^i\partial_j=\\
\fl\qquad&=-\ou{\sigma}{A}{Bi}L^i=-2\ou{\bigl[\vec{L}\!\cdot\!\vec{S}\bigr]}{A}{B}=-\ou{\bigl[\vec{J}^{\,2}-\vec{L}^2-\vec{S}^2\bigr]}{A}{B},\label{transdiracspin}
\end{eqnarray}
where $\vec{J}=\vec{L}+\vec{S}$ denotes the total angular momentum operator. We can thus rewrite the differential equation \eref{pieq} in the following diagonal form
\begin{equation}\fl\qquad
\left(\mathds{1}+\vec{J}^{\,2}-\vec{L}^2-\vec{S}^2\right)\vo{\pi}{1}{A}(\vartheta,\varphi)=
\frac{1}{2\I}\partial^k_r\ou{\big[\sigma_k\sigma^i\sigma_j\big]}{A}{B}\vou{A}{2\!}{j}{i}(\vartheta,\varphi)\vo{\pi}{0}{B}.\label{pi1eq}
\end{equation}
This equation can be solved immediately, for $\mathds{1}+\vec{J}^{\,2}-\vec{L}^2-\vec{S}^2$ has no vanishing eigenvalues\,---\,its complete spectrum is given by all integers except zero, which follows from the Clebsch\,--\,Gordan decomposition $\frac{1}{2}\otimes\ell=(\ell-\frac{1}{2})\oplus(\ell+\frac{1}{2})$ for $\ell>0$ and $\frac{1}{2}\otimes 0=\frac{1}{2}$ for $\ell=0$.

We then also know that $\vo{\pi}{1}{A}(\vartheta,\varphi)$ must be of definite parity: The leading order $\vou{A}{2}{i}{j}(\vartheta,\varphi)$ of the Ashtekar\,--\,Sen connection is required to be parity odd, yet $\partial^i_r=x^i/r$ is odd as well, hence the entire right hand side of equation \eref{pi1eq} is even. The angular momentum operators $\vec{L}$ and $\vec{S}$ commute with spatial reflections $P:\vec{x}\mapsto-\vec{x}$, hence they cannot change the parity. The next to leading order $\vo{\pi}{1}{A}(\vartheta,\varphi)$ is therefore an even function on the sphere,
\begin{equation}
\vo{\pi}{1}{A}(\vartheta,\varphi)=\vo{\pi}{1}{A}(\pi-\vartheta,\varphi+\pi).\label{oddpi}
\end{equation}
Having shown that $\vo{\pi}{1}{A}(\vartheta,\varphi)$ exists, we are left to show that the reminder $\varepsilon^A(\vec{x}):=\pi^A(\vec{x})-\vo{\pi}{0}{A}-\vo{\pi}{1}{A}(\vartheta,\varphi)/r$ in the asymptotic expansion of equation \eref{taylorpi} vanishes faster than $r^{-1}$. This is true \cite{Parker:1981uy}, and Witten's argument \cite{Wittenproof} goes as follows: By assumption, $\pi^A(\vec{x})$ solves the differential equation $\ou{D}{A}{B}\pi^B=0$, hence $\varepsilon^A$ is sourced by the spinor $j^A:=-\ou{D}{A}{B}(\vo{\pi}{0}{B}+\vo{\pi}{1}{B}/r)$. But we also know that $\vo{\pi}{0}{A}+\vo{\pi}{1}{B}/r$ was constructed so as to solve the differential equation to leading order $o(r^{-2})$. Therefore, $j^A$ must vanish as $o(r^{-2})$ for $r\rightarrow\infty$. Next, we use the Green's function of the Dirac operator to solve for $\varepsilon^A$ in terms of $j^A$. The Dirac operator $\ou{D}{A}{B}$  has the asymptotic expansion $\ou{D}{A}{B}=\ou{\sigma}{A}{Bi}\partial^i+o(r^{-2})$, hence its inverse $\ou{S}{A}{B}$ goes as
\begin{equation}
\ou{S}{A}{B}(\vec{x},\vec{y})=\frac{1}{4\pi}\frac{\ou{\sigma}{A}{Bi}(x^i-y^i)}{\left|\vec{x}-\vec{y}\right|^3}+o(r^{-2}),\qquad r\longrightarrow\infty.\label{DiracGreen}
\end{equation}
The first term represents the Green's function of the flat Dirac operator $\ou{\sigma}{A}{Bi}\delta^{ij}\partial/\partial x^j$. We now use the Green's function \eref{DiracGreen} to solve for the reminder $\varepsilon^A$ in terms of its source $j^A$, thus $\varepsilon^A(\vec{x})=\int_{\varSigma} d^3y\sqrt{\det h_{ij}(\vec{y})} \ou{S}{A}{B}(\vec{x},\vec{y})j^B(\vec{y})$. 
Since the source $j^A(\vec{x})$ vanishes as $o(r^{-2})$ for $|\vec{x}|=r\rightarrow\infty$, the integral $\int_{\varSigma} d^3y\sqrt{\det h_{ij}(\vec{y})} \ou{S}{A}{B}(\vec{x},\vec{y})j^B(\vec{y})$ is dominated by the Green's function of the flat Dirac operator $\ou{\sigma}{A}{Bi}\delta^{ij}\partial/\partial x^j$. For definiteness, assume $j^A(\vec{x})=O(r^{-2-\varepsilon})$, for $\varepsilon>0$. The leading integral $\int_{|\vec{y}|>r}d^3y\,\vec{\sigma}\!\cdot\!(\vec{x}-\vec{y})/|\vec{x}-\vec{y}|^3 j(\vec{y})$ in the \emph{far zone} $\vec{y}:|\vec{y}|>r=\vec{x}$, goes as $\int_r^\infty\di y\,{y}^{-2-\varepsilon}=r^{-1-\varepsilon}/(1+\varepsilon)$, and the leading term $\int_{\ell<\vec{y}<r}d^3y\,\vec{\sigma}\!\cdot\!(\vec{x}-\vec{y})/|\vec{x}-\vec{y}|^3 j(\vec{y})$ in the \emph{near zone} $\vec{y}:|\vec{y}|<r=\vec{x}$ falls of as $r^{-2}\int \di r\,r^{-\varepsilon}=r^{-1-\varepsilon}/(1-\varepsilon)+C'/r^2$ for $\varepsilon\neq 1$ and as $r^{-2}\int \di r\,r^{-\varepsilon}=\ln r/r^2+C/r^2$ for $\varepsilon=1$, with some short distance cut-off $\ell>0$. Thus $\varepsilon^A(\vec{x})=o(r^{-1})$, which proves our assumption \eref{taylorpi} provided the integral $\varepsilon^A(\vec{x})=\int_{\varSigma} d^3y\sqrt{\det h_{ij}(\vec{y})} \ou{S}{A}{B}(\vec{x},\vec{y})j^B(\vec{y})$ exists.

\subsection{Perturbative solution of the second spinor equation}\label{sec3.3} We can now go to the second equation \eref{witteq2}, and consider the following ansatz for its perturbative solution
\begin{equation}\fl\qquad
\omega^A(\vec{x})=\vo{\omega}{0}{A}+\ou{\sigma}{AA'}{i}x^i\,\vu{\bar\pi}{0}{A'}+\varphi^A(\vartheta,\varphi)+o(r^{0}),\qquad r\longrightarrow\infty,\label{taylorom}
\end{equation}
where the spinor $\varphi^A(\vartheta,\varphi)$ may contain all spherical harmonics $Y_{jm}(\vartheta,\varphi)$ except $Y_{00}(\vartheta,\varphi)$, which we absorb into the definition of $\vo{\omega}{0}{A}$, that is
\begin{equation}
\varphi^A(\vartheta,\varphi)=\sum_{\ell=1}^\infty\sum_{m=-\ell}^\ell\varphi^A(\ell m)Y_{\ell m}(\vartheta,\varphi).
\end{equation}
In the last section, we have studied the $r^{-1}$ expansion of $\pi^A(\vec{x})$ and have argued that there are asymptotically constant solutions $\pi^A(\vec{x})=\vo{\pi}{0}{A}+O(r^{-1}),\;r\rightarrow\infty$, which solve the first spinor equation \eref{witteq1} to leading order in $r$. We now want to show that these solutions are compatible with both the second spinor equation \eref{witteq2} and our ansatz \eref{taylorom} for the $r^{-1}$ expansion of $\omega^A(\vec{x})$. 

In doing so, we have to first show that the first three terms in \eref{taylorom} together with the leading terms in the expansion \eref{taylorpi} of $\pi^A(\vec{x})$ solve the second spinor equation \eref{witteq2} to leading orders in $r$. We thus have to solve the equation
\begin{equation}
\ou{\sigma}{A}{Ba}h^{ab}D_b\utilde{\omega}^B+3\ou{\delta}{A}{\bar A}\utilde{\bar\pi}^{A'}=o(r^{-1}),\label{trunceq}
\end{equation}
for $\utilde{\omega}^A$, where we have introduced the truncations
\begin{subalign}
\utilde{\omega}^A(\vec{x})&=\vo{\omega}{0}{A}+\ou{\sigma}{AA'}{i}x^i\,\vu{\bar\pi}{0}{A'}
+\varphi^A(\vartheta,\varphi),\label{tildeom}\\
\utilde{\pi}^A(\vec{x})&=\vo{\pi}{0}{B}+\frac{\vo{\pi}{1}{B}(\vartheta,\varphi)}{r}\label{tildepi}.
\end{subalign}\noindent
Once we have found such an $\utilde{\omega}^A$, we solve $\ou{D}{A}{B}\omega^B+3\ou{\delta}{A}{A'}\bar{\pi}^{A'}\stackrel{!}{=}0$ for the reminder $\varepsilon^A=\omega^A-\utilde{\omega}^A$ and show that $\varepsilon^A$ goes as $o(r^0)$ for $r\rightarrow\infty$.

In solving \eref{trunceq} to leading order in $r$, we can then drop all terms that vanish faster than $r^{-2}$. Going back to the asymptotic $r\rightarrow\infty$ expansions for both triad $\ou{e}{i}{a}$ and connection $\ou{A}{i}{a}$, i.e.\ equations \eref{Taylortriad} and \eref{Taylorashtekar}, we find that the leading $O(r^0)$ contributions cancel, and we are left with the $O(r^{-1})$ terms, which vanish provided
\begin{eqnarray}\nonumber\fl\qquad
\ou{\hat{\partial}}{A}{B}&\varphi^B(\vartheta,\varphi)=2\delta^{AA'}\,\vou{f}{1\!}{j}{j}(\vartheta,\varphi)\vu{\bar{\pi}}{0}{A'}+\\
\fl\qquad&-\frac{1}{2\I}\ou{\big[\sigma_l\sigma_m\sigma_n\big]}{A}{B}\partial^n_r\,\vo{A}{2\!}{ml}(\vartheta,\varphi)\delta^{BB'}\vu{\bar{\pi}}{0}{B'}
+3\delta^{AA'}\,\vu{\bar{\pi}}{1}{A'}(\vartheta,\varphi).\label{phieq1}
\end{eqnarray}
where $\ou{\hat{\partial}}{A}{B}$ denotes again the transversal component \eref{transdirac} of the three-dimensional Dirac operator \eref{Witteq0} on the spatial slice $\varSigma$. At this point, the parity conditions crucially enter the problem: As we will see in a minute, equation \eref{phieq1} can be solved for $\varphi^A(\vartheta,\varphi)$ in terms of $\bar{\pi}_{A'}$ only if the right hand side does not contain the $\ell=0$ spherical harmonic. The parity conditions guarantee this, but they are actually much stronger than that, for they imply that the entire right hand side of \eref{phieq1} is odd.

 This can be seen as follows: We multiply equation \eref{phieq1} from the left by the radial Pauli matrix $\partial^i_r\ou{\sigma}{A}{Bi}$ and use equation \eref{transdiracspin} to write the transversal Dirac operator $\ou{\hat{\partial}}{A}{B}$ in terms of the spin, angular and total angular momentum operators $\vec{S}=\vec{\sigma}/2$, $\vec{L}=-\I\vec{x}\times\partial/\partial{\vec{x}}$ and $\vec{J}=\vec{S}+\vec{L}$. 
Next, we define the source
\begin{eqnarray}\fl\qquad
\nonumber J^A&(\vartheta,\varphi)=-2\ou{\sigma}{AA'}{r}\,\vou{f}{1\!}{m}{m}(\vartheta,\varphi)\,\vu{\bar\pi}{0}{A'}+\\
\fl\qquad\nonumber &+\frac{1}{2}\left[\delta^{AA'}\vou{K}{2\!}{m}{m}(\vartheta,\varphi)-\epsilon_{lmn}\,\vo{\Gamma}{2}{lm}(\vartheta,\varphi)\,\ou{\big[\sigma_r\sigma_n\sigma_r\big]}{A}{B}\delta^{BA'}\right]\vu{\bar\pi}{0}{A'}+\\
\fl\qquad&-3\ou{\sigma}{AA'}{r}\,\vu{\bar\pi}{1}{A'}(\vartheta,\varphi),\label{sourcedef}
\end{eqnarray}
where $\ou{\sigma}{AA'}{r}=\ou{\sigma}{AA'}{i}\partial^i_r$ and $\ou{\sigma}{A}{Br}=\ou{\sigma}{A}{Bi}\partial^i_r$ are the Pauli matrices in the radial direction, and $\vou{\Gamma}{2}{i}{j}$ and $\vou{K}{2\!}{i}{j}$ are the leading $O(r^{-2})$ terms in the asymptotic $r\rightarrow\infty$ expansion of the selfdual connection
\begin{equation}
\vou{A}{}{i}{j}(r,\vartheta,\varphi)=\frac{\vou{\Gamma}{2}{i}{j}(\vartheta,\varphi)+\I\vou{K}{2\!}{i}{j}(\vartheta,\varphi)}{r^2}+o(r^{-2}).\label{OdertwoAshtekar}
\end{equation}

Having defined the source $J^A$, we can now write the defining differential equation \eref{phieq1} for $\varphi^A(\vartheta,\varphi)$ in the following diagonal form
\begin{equation}
\Big(\vec{J}^{\,2}-\vec{L}^2-\vec{S}^2\Big)\varphi^A(\vartheta,\varphi)=J^A(\vartheta,\varphi).\label{phieq2}
\end{equation}
The operator $(\vec{J}^{\,2}-\vec{L}^2-\vec{S}^2)$ can be immediately diagonalised through the Clebsch\,--\,Gordan decomposition $\frac{1}{2}\otimes \ell=(\ell-\frac{1}{2})\oplus(\ell+\frac{1}{2})$ for $\ell>0$ and $\frac{1}{2}\otimes0=\frac{1}{2}$ for $\ell=0$. The possible eigenvalues are the numbers $\Z-\{-1\}$. The single vanishing eigenvalue comes from $\ell=0$. We can thus solve equation \eref{phieq2} for $\varphi^A(\vartheta,\varphi)$ in terms of $J^A(\vartheta,\varphi)$ only for a source $J^A(\vartheta,\varphi)$ that does not contain the $\ell=0$ spherical harmonic\,---\,it must admit the expansion
\begin{equation}
J^A(\vartheta,\varphi)=\sum_{\ell=1}^\infty\sum_{m=-\ell}^\ell J^A(\ell m) Y_{\ell m}(\vartheta,\varphi).
\end{equation}
This is true thanks to the parity conditions on both the triad and connection: The leading order of the metric perturbation $\vou{f}{1}{i}{j}(\vartheta,\varphi)$ is even, and $\vou{A}{2\!}{i}{j}(\vartheta,\varphi)$ is odd. The outwardly oriented normal vector $\partial^i_r=x^i/r$ is also odd, hence the radial Pauli matrix $\sigma_r=\sigma_i\partial^i_r$ is odd as well. Going back to the definition \eref{sourcedef} for the source $J^A(\vartheta,\varphi)$, we thus see that $J^A(\vartheta,\varphi)$ is also odd. The spherical harmonics $Y_{\ell m}$ are of alternating parity, $Y_{\ell m}(\vartheta,\varphi)=(-1)^{2\ell}Y_{\ell m}(\pi-\vartheta,\varphi+\pi)$, and the source $J^A(\vartheta,\varphi)$ does, therefore, not contain the $\ell=0$ spherical harmonic, which implies in turn that we can solve \eref{phieq2} for $\varphi^A(\vartheta,\varphi)$ in terms of $J^A(\vartheta,\varphi)$. The solution is unique, because we have already absorbed the $\ell=0$ mode of $\varphi^A(\vartheta,\varphi)$ into the definition of $\vo{\omega}{0}{A}$. Furthermore, with $J^A(\vartheta,\varphi)$ being odd, $\varphi^A(\vartheta,\varphi)$ is also odd, because $P:\vec{x}\rightarrow-\vec{x}$ commutes with the operator $\vec{J}^{\,2}-\vec{L}^2-\vec{S}^2$.

In summary, what we have shown so far is this: The leading terms in our ansatz \eref{taylorom} can be always chosen so as to solve the differential equation \eref{witteq2} to order $o(r^{-2})$. The solution for $\varphi^A(\vartheta,\varphi)$ is unique, and parity odd provided the parity conditions on triad and connection are satisfied. The only ambiguity is in $\vo{\omega}{0}{A}$, which is a free integration constant. What we have to show now is that the remainder $\varepsilon^A(\vec{x}):=\omega^A(\vec{x})-\utilde{\omega}^A(\vec{x})$, with $\utilde{\omega}^A(\vec{x})$ defined as in \eref{tildeom}, exists and vanishes faster than $r^0$. We follow the strategy \cite{Wittenproof} of Witten, as discussed in the last section. We insert $\varepsilon^A(\vec{x})$ into the three-dimensional Dirac equation and require $\utilde{\omega}^A(\vec{x})+{\varepsilon}^A(\vec{x})$ be a solution of the second spinor equation \eref{witteq2}. Now, $\utilde{\omega}^A$ solves the truncation \eref{trunceq} to order $o(r^{-1})$, but $\omega^A$ is assumed to be an exact solution, hence the remainder $\varepsilon^A$ is subject to the equation
\begin{equation}\fl\qquad
\ou{\sigma}{A}{Ba}h^{ab}D_b\varepsilon^B=3\delta^{AA'}\big(\bar{\pi}_{A'}-\bar{\utilde{\pi}}_{A'}\big)+o(r^{-1})=o(r^{-1}),\quad r\rightarrow\infty,
\end{equation}
where we used the asymptotic $r\rightarrow\infty$ expansion of both $\pi^A(\vec{x})$ and $\utilde{\omega}^A(\vec{x})$ as in \eref{taylorpi}, \eref{trunceq} and \eref{tildepi}. 

We thus see that $\ou{D}{A}{B}\varepsilon^B$ is sourced by a spinor $j^A(\vec{x})$, which falls of as $o(r^{-1})$. For simplicity let us assume\footnote{The argument also works for $j^A(\vec{x})=O(\ln r/r^{\varepsilon})$ and any $\varepsilon>0$.} that $j^A(\vec{x}$) is of order $O(r^{-1-\varepsilon})$, for some $\varepsilon>0$. Going back to the 
Green's function \eref{DiracGreen}, we can thus solve for $\varepsilon^A(\vec{x})$ in terms of the source $j^A(\vec{x})$. We then have $\varepsilon^A(\vec{x})=\int_{\varSigma} d^3y\sqrt{\det h_{ij}(\vec{y})}\,\ou{S}{A}{B}(\vec{x},\vec{y})j^B(\vec{y})$, which vanishes as $O(r^{-\varepsilon})$. This can be seen as follows: Going back to the $r^{-1}$ expansion of the Green's function, we see that the asymptotic $r=|\vec{x}|\rightarrow\infty$ behaviour of the integral is dominated by $\vou{S}{2}{A}{B}(\vec{x},\vec{y})=1/(4\pi)\ou{\sigma}{A}{Bi}(x^i-y^i)/|\vec{x}-\vec{y}|^3$, which is the Green's function of the free Dirac operator $\ou{\sigma}{A}{Bi}\partial/\partial x^i$. In the \emph{near zone} $\ell<|\vec{y}|<|\vec{x}|=r$, the dominant contribution is given by the integral $\frac{1}{r^2}\int_\ell^r\di y\,{y}^{1-\varepsilon}=r^{-\varepsilon}/({2-\varepsilon})+C/r^2$ for $\varepsilon\neq 2$ and $\frac{1}{r^2}\int_\ell^r\di y\,{y}^{-1}=\ln r/r^2+C/r^2$ for $\varepsilon=2$, with some short distance cut-off $\ell>0$. In the \emph{far zone} $\vec{y}>|\vec{x}|=r$, on the other hand, the dominant contribution comes from the integral $\int_r^\infty \di y\,{y}^{-1-\varepsilon}=\varepsilon^{-1} r^{-\varepsilon}$. Hence $\varepsilon^A(\vec{x})=o(r^0)$, provided the integral converges, in which case the asymptotic expansion \eref{taylorom} exists and provides a solution of the second spinor equation \eref{witteq2}.

\section{\textsc{ADM} charges and the evaluation of the boundary integral}\label{sec4}
We are now ready to evaluate the charge $Q_{\pi,\omega}$ at spacelike infinity. 
The perturbative evaluation of the boundary integral $Q_{\pi,\omega}$ greatly simplifies in the language of differential forms. We go back to the definition \eref{Qcharge} of $Q_{\pi,\omega}$ in terms of $\pi^A$ and $\omega^A$ and insert the first spinor equation \eref{witteq1} in the form of $\hat{r}^aD_a\pi^A=\I\hat{r}^b\uo{\varepsilon}{b}{ac}\ou{\sigma}{A}{Ba}D_c\pi^B$, with $\hat{r}^a$ denoting the outwardly oriented unit normal of $S^2_r$ with respect to the physical three-metric $h_{ab}=g_{ab}+n_a n_b$. We can then write $Q_{\pi,\omega}$ as the integral
\begin{equation}
Q_{\pi,\omega}=-\I\lim_{r\rightarrow\infty}\int_{S^2_r}\bar{\omega}^{\bar A}\sigma_{AA'}\wedge D\pi^A,\label{Qcharge2}
\end{equation}
where $\sigma_{AA'}$ are the soldering forms $\sigma_{AA'}=\sigma_{AA'\alpha}e^\alpha$ for the tetrad $e^\alpha$. To use tetrads, rather than triads (as we did in the last section), may seem a little odd, since we are always staying on a given three-dimensional initial hypersurface $\varSigma$, but this trick is useful for us, because it keeps the formalism manifestly covariant. We should then also extend our asymptotic Cartesian coordinates $\{x^i\}$ into an asymptotic rest frame $\{x^0,x^i\}$ near spacelike infinity, such that $\varSigma$ turns (for large distance) into the $x^0=0$ hypersurface. We then also assume that the \emph{tetrad} $e^\alpha$ admits the asymptotic $r^{-1}$ expansion around $\varSigma$
\begin{equation}
e^\alpha=\di x^\alpha+f^\alpha=D(r\partial^\alpha_r)+\tilde{f}^\alpha,\label{taylortetra}
\end{equation}
where the vector valued one-forms $f^\alpha$ and $\tilde{f}^\alpha$ are of order $O(r^0)$, hence their component functions $\ou{f}{\alpha}{\mu}:=\ou{f}{\alpha}{a}\partial^a_\mu$ with respect to the asymptotic rest frame $\{x^\mu\}$ are of order $O(r^{-1})$.  This agrees with the asymptotic expansion of the triad \eref{taylortriad}, which is the pull-back of $e^\alpha$ to $\varSigma$ (we can always choose an internal Lorentz gauge for which $n^\alpha=\ou{e}{\alpha}{a}n^a=\delta^\alpha_0$, in which case the pull-back of $\delta^i_\mu e^\mu$ to the spatial hypersurface $\varSigma$ returns the triad $e^i$). 

In defining \eref{taylortetra}, we have extended the radial coordinate $r=\sqrt{\delta_{ij}x^ix^j}$ on $\varSigma:x^0=0$ into the hyperbolic coordinate $r=\sqrt{\eta_{\mu\nu}x^\mu x^\nu}$ around a four-dimensional neighbourhood of spacelike infinity.   The corresponding radial vector field is $\partial^a_r\in TM$, which becomes $\partial^\alpha_r:={e}{}^{\alpha}{}_{a}\partial^a_r$ in internal space. The difference  between $f^\alpha$ and $\tilde{f}^\alpha$ is measured by the spin rotation coefficients through $f^\alpha-\tilde{f}^{\alpha}=r\ou{A}{\alpha}{\beta}\partial^\beta_r+D(r\ou{f}{\alpha}{a}\partial^a_r)$, where $\ou{A}{\alpha}{\beta}$ denotes the torsionless spin connection in a neighbourhood of $\varSigma$. Its pullback to $\varSigma$ admits the asymptotic expansion \eref{Taylorashtekar}. The leading orders $\vou{\tilde{f}}{1}{\mu}{\nu}$ and $\vou{A}{2}{\alpha}{\beta\mu}$ of the component functions of tetrad and connection are assumed to have definite parity: $\vou{\tilde{f}}{1}{\mu}{\nu}$ is even and  $\vou{A}{2}{\alpha}{\beta\mu}$ is odd, in accordance with the parity conditions in the rest of this paper.

We then also know from \eref{taylorpi} and \eref{taylorom} that the spinors $\pi^A$ and $\omega^A$ admit the asymptotic expansions $\pi^A=\vo{\pi}{0}{A}+\vo{\pi}{1}{A}/r+o(r^{-1})$ respectively $\omega^A=\vo{\omega}{0}{A}+\phi^A+r\ou{\sigma}{AA'}{\alpha}\partial^\alpha_r{\bar\pi}_{A'}+o(r^{0})$ on $\varSigma$, where $\phi^A$ differs from $\varphi^A$ (as in equation \eref{taylorom}) by the term $\ou{\sigma}{AA'}{\mu}\vu{\bar\pi}{0}{A'}\vou{f}{1}{\mu}{a}\partial^a_r+\ou{\sigma}{AA'}{\alpha}\partial^\alpha_r\vu{\bar{\pi}}{1}{A'}=\varphi^A-\phi^A$, which is odd, because $\vo{\pi}{1}{A}$ is even whereas $\varphi^A$ is odd.  The spinor $\phi^A(\vartheta,\varphi)$ is therefore an odd function on the sphere, and we will crucially need this in the following.

With these preparations in mind, we can now insert the asymptotic $r\rightarrow\infty$ expansion of the spinors \eref{taylorpi} and \eref{taylorom} into the expression for the asymptotic charge \eref{Qcharge2}. In evaluating this integral for $r\rightarrow\infty$, we drop all terms that fall of as $r^{-1}$ or faster and find
\begin{eqnarray}\nonumber\fl\qquad
Q_{\pi,\omega}=-\I\lim_{r\rightarrow\infty}\int_{S^2_r}&\left[\left(\vo{\bar\omega}{0}{A'}+\bar{\phi}^{A'}\right)\sigma_{AA'a} D(r\partial^a_r)\wedge D\pi^A+\right.\\
\fl\qquad&\left.+r\ou{\sigma}{BA'}{\beta}\sigma_{AA'\alpha}\partial^\beta_r\left(D(r\partial^\alpha_r)+\tilde{f}^\alpha\right)\wedge D\pi^A\,\pi_{B}\right].\label{Qint1}
\end{eqnarray}
We study both terms separately: In the first line of  equation \eref{Qint1}, we use Stokes's theorem on $S^2_r$, which is closed, and perform a partial integration. The soldering forms are covariantly constant\footnote{$D_a$ acts on all indices: on spinor indices $A,B,\dots$ and $A',B',\dots$ through the torsionless spin connection, on tangent indices $a,b,c,\dots$ through the Christoffel symbols for the spatial metric.}, i.e.\ $D_a{\sigma}{}^{AA'}{}_{b}=0$, hence
\begin{eqnarray}\fl\nonumber
\mbox{\qquad eq.\,\eref{Qint1}, $1^{\rm st}$ line}=-\I&\lim_{r\rightarrow\infty}\int_{S^2_r}\left[-r\sigma_{AA'a}\partial^a_r\,D\bar{\phi}^{A'}\wedge D\pi^A+\right.\\
&\left.+r\left(\bar{\epsilon}_{A'B'}F_{AB}\right)\ou{\sigma}{AA'}{a}\partial^a_r\,\vo{\pi}{0}{B}\left(\vo{\bar\omega}{0}{B'}+\bar{\phi}^{B'}\right)\right],\label{Qint2}
\end{eqnarray}
where we dropped already all terms in the integral that fall of as $r^{-1}$ or faster, and used the definition of the field strength as the square of the exterior covariant derivative, i.e.\ $D^2\pi^A=\ou{F}{A}{B}\pi^B$. No terms containing the differential $\di r$ appear, because the two-sphere $S^2_r$ is meant to be a level surface of constant $r$. The first line of \eref{Qint2} vanishes due to parity: $\partial^a_\mu D_a\phi^A$ is $O(r^{-1})$ even and $\partial^a_\mu D_a\pi^A$ is $O(r^{-2})$ odd. Only the second line of \eref{Qint2} can contribute to the integral. It contains $-\epsilon_{A'B'}F_{AB}$, which is the selfdual component $(\ast+\I)F_{\alpha\beta}$ of the $SO(1,3)$ curvature two-form  $\ou{F}{\alpha}{\beta}=\di\ou{A}{\alpha}{\beta}+\ou{A}{\alpha}{\mu}\wedge\ou{A}{\mu}{\beta}$ in spinor notation. The connection is torsionless $De^\alpha=\di e^\alpha+\ou{A}{\alpha}{\beta}\wedge e^\beta$, which implies that $F_{\alpha\beta ab}=\uo{e}{\alpha}{c}\uo{e}{\beta}{d}R_{cdab}$, with $R_{abcd}$ denoting the Riemann curvature tensor. At spacelike infinity there should be no matter, hence $R_{abcd}=C_{abcd}$, with $C_{abcd}$ denoting the Weyl tensor, which can be split into its electric and magnetic components
\begin{subalign}
E_{ab}&={}^{\phantom\ast} C_{acbd}\partial^c_r\partial^d_r,\label{Efield}\\
B_{ab}&={}^\ast C_{acbd}\partial^c_r\partial^d_r=\frac{1}{2}\ou{\varepsilon}{ef}{ac}C_{efbd}\partial^c_r\partial^d_r.\label{Bfield}
\end{subalign}\noindent
We insert this decomposition into \eref{Qint2}. After some straight-forward algebraic manipulations we are left with
\begin{eqnarray}\fl
\mbox{\quad eq.\,\eref{Qint1}, $1^{\rm st}$ line}=\frac{1}{2}\lim_{r\rightarrow\infty}\int_{S^2_r}d^2v\,r\left(E_{ab}-\I B_{ab}\right) n^a\uo{\sigma}{BB'}{b}\,\vo{\pi}{0}{B}\big(\vo{\bar\omega}{0}{B'}+\bar{\phi}^{B'}\big),\label{ElectricInt}
\end{eqnarray}
where $d^2v=\frac{1}{2}\tilde{\eta}^{bc}\hat{r}^a\varepsilon_{abc}$ is the canonical volume element on the sphere ($\hat{r}^a$ is the outwardly oriented unit normal of $S^2_r$ in $\varSigma$).

Let us now turn to the second line of equation \eref{Qint1}. The first simplification is found by decomposing $D\pi^A\pi_B$ into its $SL(2,\C)$ irreducible components $\pi_AD\pi^A$ and $\pi^{(A}D\pi^{B)}$ corresponding to spin $0$ and spin $1$. This is achieved through the generalised Pauli identity for the soldering forms $\ou{\sigma}{AA'}{\alpha}$. We have
\begin{equation}\fl\quad
\ou{\sigma}{AC'}{\alpha}\sigma_{BC'\beta}=-\delta^A_B\eta_{\alpha\beta}-2\ou{\Sigma}{A}{B\alpha\beta},\quad
\ou{\Sigma}{AB}{\alpha\beta}:=\frac{1}{2}\ou{\sigma}{A}{C'[\alpha}\ou{\bar\sigma}{C'B}{\beta]},\label{fourdimPauli}
\end{equation}
with $\ou{\Sigma}{A}{B\alpha\beta}$ forming the  selfdual basis\footnote{The electric component of $\ou{\Sigma}{A}{Bab}$  is $\ou{\Sigma}{A}{Bba}n^b=\frac{1}{2}\ou{\sigma}{A}{Ba}$, its magnetic component is $\frac{1}{2}\uo{\varepsilon}{a}{cb}\ou{\Sigma}{A}{Bbc}=\frac{1}{2\I}\ou{\sigma}{A}{Ba}$, with $\ou{\sigma}{A}{Ba}$ denoting the Pauli matrices \eref{pmatrx} on the spatial slice $\varSigma$.} in the Lie algebra $\mathfrak{sl}(2,\C)$. 
Splitting $D\pi^A\pi_B$ into its irreducible components, we can thus bring the second line of equation \eref{Qint1} into the form
\begin{eqnarray}\fl
\mbox{\quad eq.\,\eref{Qint1}, $2^{\rm nd}$ line}=-\I\lim_{r\rightarrow\infty}\int_{S^2_r}&\left[r\Sigma_{AB\alpha\beta}\partial^\alpha_r e^\beta\wedge D(\pi^A\pi^B)-r\partial^\alpha_r e_\alpha\wedge D\pi^A\pi_A\right].\label{Qint3}
\end{eqnarray}
The second term of \eref{Qint3} vanishes again due to parity: The components of the one-form $\partial^\alpha_r e_\alpha$ in the asymptotic frame $\{x^\mu\}$ are $O(r^{-1})$ and odd, and the components of $D\pi^A$ are of order $O(r^{-2})$ but also odd. The coordinate volume $\di\vartheta\wedge\di\varphi$ is odd as well, hence the second term vanishes as $r\rightarrow\infty$. We are thus left with just the first term. Performing a partial integration in $S^2_r$ yields
\begin{eqnarray}
\mbox{eq.\,\eref{Qint1}, $2^{\rm nd}$ line}=-\I\lim_{r\rightarrow\infty}\int_{S^2_r}&r\Sigma_{AB\alpha\beta}D\partial^\alpha_r \wedge e^\beta\,\pi^A\pi^B,\label{Qint4}
\end{eqnarray}
where we used the vanishing of torsion $De^\alpha=0$ and the covariant constancy of $\ou{\Sigma}{A}{B\alpha\beta}$. To simplify this term, we go back to \eref{taylortetra} and write $D(r\partial^\alpha_r)$ as the difference between the tetrad $e^\alpha$ and $\tilde{f}^\alpha$. Thus
\begin{eqnarray}\fl\nonumber\quad
&\int_{S^2_r}r\Sigma_{AB\alpha\beta}D\partial^\alpha_r \wedge e^\beta\,\pi^A\pi^B=\int_{S^2_r}\Sigma_{AB\alpha\beta}(e^\alpha-\tilde{f}^\alpha) \wedge e^\beta\,\pi^A\pi^B=\\
\fl&\qquad=\int_{S^2_r}\Sigma_{AB\alpha\beta}e^\alpha \wedge e^\beta\,\pi^A\pi^B-\int_{S^2_r}\left[\Sigma_{AB\alpha\beta}\tilde{f}^\alpha\wedge\left(rD\partial^\beta_r+\tilde{f}^\beta\right)\pi^A\pi^B\right].\label{Qint5}
\end{eqnarray}
The first term in the second line of \eref{Qint5} (this is Bramson's integrand \cite{BramsonSpin} for the intrinsic spin of a gravitational system) vanishes thanks to the defining differential equation \eref{witteq1}. This can be seen by applying Stokes's theorem to write the surface integral over $S^2_r$ as a vanishing volume integral over the bulk. If $B_r\subset\varSigma$ denotes the three-ball bounded by $S^2_r=\partial B_r$, we have 
\begin{eqnarray}\nonumber
\int_{\partial B_r}\!\Sigma_{AB\alpha\beta}e^\alpha& \wedge e^\beta\,\pi^A\pi^B=2\int_{B_r}\!\Sigma_{AB\alpha\beta}e^\alpha \wedge e^\beta\wedge D\pi^A\pi^B=\\
&=\I\int_{B_r}\!\!d^3 v\sigma_{ABa}h^{ab}D_b\pi^A\pi^B=0.\label{vanshbulk}
\end{eqnarray}
In going from the first line to the second line of \eref{vanshbulk}, we used the identity $\varepsilon^{abc}\Sigma_{AB\beta\gamma}\ou{e}{\beta}{b}\ou{e}{\gamma}{c}=\I\uo{\sigma}{AB}{a}$, which follows from the definition \eref{pmatrx} of the Pauli matrices $\ou{\sigma}{A}{Ba}$ on $\varSigma$. The second line vanishes identically, because $\pi^A$ is a solution of its defining differential equation \eref{witteq1}.
Since $\int_{S^2_r}\Sigma_{AB}\pi^A\pi^B$ thus vanishes, we are left with the last term in equation \eref{Qint5}. We perform a partial integration in $S^2_r$, and move the exterior covariant derivative $D$ away from $\partial^\alpha_r$. In the limit of $r\rightarrow\infty$ all but one term vanish\footnote{The $O(r^0)$ vanishing terms are $\Sigma_{AB\alpha\beta}\tilde{f}^\alpha\wedge \tilde{f}^\beta\vo{\pi}{0}{A}\vo{\pi}{0}{B}$ and $2\Sigma_{AB\alpha\beta}\partial^\alpha_r\tilde{f}^\beta \wedge(\di\vo{\pi}{1}{A}+\vou{A}{2}{A}{C}\vo{\pi}{0}{C})\vo{\pi}{0}{B}$, which are odd, since both $\vo{\pi}{1}{A}(\vartheta,\varphi)$ and $\vou{f}{1}{\alpha}{\mu}(\vartheta,\varphi)$ are even, but the integration measure $\di\vartheta\wedge\di\varphi$ and the leading $O(r^{-2})$ connection coefficient $\vou{A}{2}{A}{B}(\vartheta,\varphi)$ are parity odd. } due to parity, and we are arrive at
\begin{eqnarray}\fl\quad
&\lim_{r\rightarrow\infty}\int_{S^2_r}\left[\Sigma_{AB\alpha\beta}\tilde{f}^\alpha\wedge\left(rD\partial^\beta_r+\tilde{f}^\beta\right)\pi^A\pi^B\right]=\lim_{r\rightarrow\infty}\int_{S^2_r}\left[r\Sigma_{AB\alpha\beta}D\tilde{f}^\alpha\partial^\beta_r\pi^A\pi^B\right].\label{Qint6}
\end{eqnarray}
The right hand side crucially contains the two-form $D\tilde{f}^\alpha$, which is proportional to the magnetic part of the Weyl tensor. Indeed, we have
\begin{eqnarray}\fl\quad
\tilde{\eta}^{ab}D_a\ou{\tilde{f}}{\alpha}{b}&=\tilde{\eta}^{ab}D_a\left({e}{}^{\alpha}{}_{b}-D_b(r\partial^\alpha_r)\right)=-\frac{r}{2}\tilde{\eta}^{ab}\ou{F}{\alpha}{\beta ab}\partial^\beta_r=-r\,d^2v\,e^{\alpha a}B_{ab}\partial^b_r,
\end{eqnarray}
with $\tilde{\eta}^{ab}$ denoting the metric independent Levi-Civita volume-density on $S^2_r$.  Going back over equations \eref{vanshbulk} and \eref{Qint5} to equation \eref{Qint4}, we have thus shown that
\begin{eqnarray}\fl
\mbox{\quad eq.\,\eref{Qint1}, $2^{\rm nd}$ line}=-\I\lim_{r\rightarrow\infty}\int_{S^2_r}d^2v\,r^2\,B_{ab}\,n^a\,\ou{\Sigma}{ABb}{c}\,\partial^c_r\,\pi_A\pi_B,\label{MagneticInt}
\end{eqnarray}
provided the parity conditions are satisfied. We combine this result with equation \eref{ElectricInt} for the first line of \eref{Qint1}. This leaves us with an integral over the electric and magnetic components (\ref{Efield}, \ref{Bfield}) of the Weyl tensor at spacelike infinity, in fact
\begin{eqnarray}\fl\qquad\nonumber
Q_{\pi,\omega}=+\frac{1}{2}\lim_{r\rightarrow\infty}\int_{S^2_r}d^2v\,r\left(E_{ab}-\I B_{ab}\right) n^a\uo{\sigma}{BB'}{b}\,\vo{\pi}{0}{B}\big(\vo{\bar\omega}{0}{B'}+\bar{\phi}^{B'}\big)+\\
-\I\lim_{r\rightarrow\infty}\int_{S^2_r}d^2v\,r^2\,B_{ab}\,n^a\,\ou{\Sigma}{ABb}{c}\,\partial^c_r\,\pi_A\pi_{B}.\label{Qint8}
\end{eqnarray}
The parity conditions further simplify this integral: The leading oder of the metric perturbation is $O(r^{-1})$ even, and the connection coefficients are $O(r^{-2})$ odd. This is essentially the same as to say that the leading $r^{-3}$ order of the Weyl tensor contains only electric components and is parity even. This in turn implies that the the spinor $\phi^A(\vartheta,\varphi)$, which we have shown to be odd\footnote{Which followed from \eref{phieq2} and the parity conditions on metric and connection.}, drops out of the final expression
\begin{eqnarray}\fl\qquad\nonumber
Q_{\pi,\omega}=\frac{1}{2}\lim_{r\rightarrow\infty}\int_{S^2_r}d^2\Omega\,r^3\,n^a\uo{E}{a}{b}\,\ou{\sigma}{BB'}{b}\,\vu{\pi}{0}{B}\vu{\bar\omega}{0}{B'}+\\
-\I\lim_{r\rightarrow\infty}\int_{S^2_r}d^2\Omega\,r^4\,n^a\uo{B}{a}{b}\,\ou{\Sigma}{AB}{bc}\,\partial^c_r\,\vu{\pi}{0}{A}\vu{\pi}{0}{B},\label{Qint7}
\end{eqnarray}
where we replaced the $O(r^2)$ metrical volume element $d^2v=\frac{1}{2}\tilde{\eta}^{bc}\hat{r}^a\varepsilon_{abc}$ in the limit of $r\rightarrow\infty$ by the fiducial volume element $d^2\Omega=\di\hspace{-0.08em}\cos\vartheta\wedge\di\varphi=\lim_{r\rightarrow\infty}r^{-2}d^2v$ on the two-sphere at spacelike infinity.

That $\phi^A(\vartheta,\varphi)$ falls out of the asymptotic charge \eref{Qint7} through the parity conditions represents the removal of the super-translation ambiguity from the definition of the angular momentum at spacelike infinity. Indeed $\phi^A(\vartheta,\varphi)$ transforms non-trivially under super-translations. We will come back to this point below. 

Equation \eref{Qint7} can be written as a linear combination of all ten Poincaré charges at spacelike infinity\,---\,of both the \textsc{ADM} four-momentum $P_\alpha$ and the relativistic angular momentum $M_{\alpha\beta}$. Indeed
\begin{equation}
Q_{\pi,\omega}=4\pi G\left(P^\alpha\ou{\sigma}{AA'}{\alpha}\vu{\pi}{0}{A}\vu{\bar\omega}{0}{A'}-M^{\alpha\beta}\ou{\Sigma}{AB}{\alpha\beta}\vu{\pi}{0}{A}\vu{\pi}{0}{B}\right),\label{QADM}
\end{equation}
where we have used the definition of the \textsc{ADM} charges $P_\alpha$ and $M_{\alpha\beta}$ as the surface integrals\begin{subalign}
P^\alpha&=\frac{1}{8\pi G}\lim_{r\rightarrow\infty}\int_{S^2_r}d^2\Omega\,r^3\,n^a\uo{E}{a}{b}\,\di x^\alpha_b\label{ADMen},\\
M^{\alpha\beta}&=\frac{1}{8\pi G}\lim_{r\rightarrow\infty}\int_{S^2_r} d^2\Omega\,r^4\,n^a\uo{B}{a}{b}\,\partial^c_r\,\ou{\varepsilon}{de}{bc}\,\di x^\alpha_d\,\di x^\beta_e,\label{ADMspin}
\end{subalign}\noindent
over the electric and magnetic components of the Weyl tensor as introduced by Ashtekar and Hansen \cite{Ashtekar:1978zz}. The notation may need to be clarified again: The fiducial tetrad $\di x^\alpha_a$ refers to the asymptotic inertial coordinate system and $\ou{\Sigma}{AB}{\alpha\beta}$ are the selfdual generators \eref{fourdimPauli} of $\mathfrak{sl}(2,\C)$. 

\paragraph{} Finally, we can write the charge \eref{QADM} as a three-dimensional volume integral over $\varSigma$, thus providing a quasi-local three-density for the \textsc{ADM} angular momentum and centre of mass. We take $Q_{\pi,\omega}$ in the form of \eref{Qcharge}, and apply Gauss's divergence theorem. This turns the surface integral over $S^2_r$ for $r\rightarrow\infty$ into the volume integral 
\begin{equation}\fl\qquad
Q_{\pi,\omega}=\int_\varSigma d^3v\left[h^{ab}\delta_{BB'}D_a\bar{\omega}^{B'}D_b\pi^B+\bar{\omega}_{A'}h^{ab}D_a\left(\uo{h}{b}{c}\delta_{AA'}D_c\pi^A\right)\right],\label{Qvol1}
\end{equation}
 over the spatial slice $\varSigma$, with $d^3v=\frac{1}{3!}\tilde{\eta}^{abc}\varepsilon_{abc}$ denoting the canonical volume element. From \eref{witteq1}, we also know that $\ou{\sigma}{A}{Ba}D^a(\ou{\sigma}{B}{Cb}D^b\pi^C)=0$. Employing the Pauli identity \eref{paulident} and the definition of the curvature tensor, we can then show that the second term in \eref{Qvol1} depends only on the Ricci tensor $R_{ab}$, but not on the derivatives of $\pi^A$. In fact
\begin{equation}
 h^{ab}D_a\left(\uo{h}{b}{c}\delta_{AA'}D_c\pi^A\right)=-\frac{1}{2}\Big(R_{ab}-\frac{1}{2}g_{ab}R\Big)n^a\uo{\sigma}{AA'}{b}\pi^A.
\end{equation}
We can then finally insert the Einstein equations for a given stress-energy tensor $T_{ab}$, and arrive at the following quasi-local expression for the \textsc{ADM} charge \eref{QADM}

\begin{equation}
Q_{\pi,\omega}=\int_{\varSigma}d^3v\left[h^{ab}\delta_{AA'}D_a\bar{\omega}^{A'}D_b\pi^A-4\pi G\,T_{ab}n^a\uo{\sigma}{BB'}{b}\pi^B\bar{\omega}^{B'}\right].\label{Qvol2}
\end{equation}
This expression, which is a version of Sparling's equation \cite{penroserindler2}, should be compared with the quasi-local expression of the \textsc{ADM} linear momentum due to Witten \cite{Wittenproof}. Indeed $Q_{\pi,\omega}$, has the very same structure. The first term of \eref{Qvol2} defines a quasi-local density for the selfdual component of the gravitational angular momentum and centre of mass shifted by the linear momentum, the second term represents the contribution from the matter fields. 


\paragraph{How does $Q_{\pi,\omega}$ transform under asymptotic isometries?} First of all, $Q_{\pi,\omega}$ is clearly invariant under proper orthochronous Lorentz transformations $x^\mu\!\longrightarrow\!\ou{\Lambda}{\mu}{\nu}x^\nu,\,\Lambda\in L^\uparrow_+$ provided the spinors transform under the corresponding $SL(2,\C)$ transformation as well, i.e.\ $\vo{\pi}{0}{A}\!\longrightarrow\!\ou{g}{A}{B}\vo{\pi}{0}{B}$ and $\vo{\omega}{0}{A}\!\longrightarrow\!\ou{g}{A}{B}\vo{\omega}{0}{B}$, with $g:\ou{\Lambda}{\alpha}{\beta}=\ou{g}{A}{B}\ou{\bar{g}}{A'}{B'}$. What about translations? In Minkowski space, the relativistic angular momentum of a particle with four-momentum $P^\alpha$ and position $X^\alpha$ is given by the tensor $M^{\alpha\beta}=-2X^{[\alpha}P^{\beta]}$, which transforms under linear translations $X^\mu\!\longrightarrow \!X^\mu+T^\mu$ as $M^{\alpha\beta}\!\longrightarrow\!M^{\alpha\beta}-2T^{[\alpha} P^{\beta]}$. The \textsc{ADM} linear momentum and angular momentum transform in exactly the same way \cite{Szabados_Poincare}. On the other hand, $\vo{\pi}{0}{A}$ is translational invariant, which follows directly from the asymptotic $r\rightarrow\infty$ expansion \eref{taylorpi} of $\pi^A$, but $\vo{\omega}{0}{A}$ is not. Going back to the asymptotic expansion \eref{taylorom} for $\omega^A$, we find, in fact $\vo{\omega}{0}{A}\!\longrightarrow\! \vo{\omega}{0}{A}-\ou{\sigma}{AA'}{\mu}T^\mu\vu{\bar\pi}{0}{A'}$, which cancels the term $T^{[\alpha} P^{\beta]}$ coming from the transformation $M^{\alpha\beta}\!\longrightarrow\!M^{\alpha\beta}-2T^{[\alpha} P^{\beta]}$ of the \textsc{ADM} angular momentum. The definition of $Q_{\pi,\omega}$ is therefore invariant under all asymptotic Poincaré transformations. This is all not very surprising, in fact it should have been expected: We have defined $Q_{\pi,\omega}$ as a two-dimensional surface integral \eref{Qcharge} over the spinors $(\bar{\pi}_{A'},\omega^A)$ that are solutions of a particular system of linear differential equations (\ref{witteq1}, \ref{witteq2}) on $\varSigma$. The entire construction of $Q_{\pi,\omega}$ is coordinate invariant, and it is only through the asymptotic $r\rightarrow\infty$ expansion (\ref{taylorpi}, \ref{taylorom}) and the subsequent perturbative evaluation of the boundary integral from \eref{Qint1} to \eref{QADM} that an explicit choice of coordinates ever enters the problem. The very value of $Q_{\pi,\omega}$ for a given solution $(\bar{\pi}_{A'},\omega^A)$ of the defining differential equations \eref{witteq1} and \eref{witteq2} cannot depend on the asymptotic rest frame around spacelike infinity. 

By the same argument, we infer how the leading $O(r^0)$ terms $\vo{\pi}{0}{A}$, $\vo{\omega}{0}{A}$ and $\varphi^A(\vartheta,\varphi)$ in (\ref{taylorpi}, \ref{taylorom}) transform under the remaining asymptotic isometries\footnote{Working on a fixed initial hypersurface $\varSigma$, we should think of these isometries as \emph{passive} coordinate transformations $x^\mu\longrightarrow\tilde{x}^\mu=\tilde{x}^\mu(x)$ that do not deform the initial hypersurface $\varSigma$, but only change the asymptotic coordinate representation of $\varSigma$ in $\R^4$}, which are \cite{AshtekarLogAmbi, Ashtekar:1978zz} even and odd super-translations $x^\mu\longrightarrow x^\mu+S^\mu(\vartheta,\varphi)$ and logarithmic translations. The charge $Q_{\pi,\omega}$ is invariant under all these isometries, but the leading order coefficients $\vo{\pi}{0}{A}$, $\vo{\omega}{0}{A}$ and $\varphi^{A}(\vartheta,\varphi)$ in the $r\rightarrow\infty$ expansion of the spinors are not. For logarithmic translations $x^\mu\longrightarrow x^\mu+C^\mu\ln r$ ($C^\mu$ is a constant function on the sphere) the argument is more subtle, because these transformations actually change the asymptotic expansion \eref{taylorom} of $\omega^A$, and generate a term that grows as $O(\ln r)$ with $r\rightarrow\infty$.

\section{Quasi-local charges and conservation laws}\label{sec5}
So far, we have only considered gravitational charges, which are defined at spacelike infinity. This is a mathematical idealisation, any actual measurement is taken at large but finite distance from the source. We should therefore rather study \emph{quasi-local charges}, and we thus close this paper with an outlook and discussion for how to compute and derive conservation laws for the quasi-local quantities
\begin{subalign}
Q_{\pi,\omega}[S]&=-\I\int_S\bar\omega^{A'}\sigma_{AA'}\wedge D\pi^A,\label{Qdefvar}\\
E_\pi[S]&=-\I\int_S\bar\pi^{A'}\sigma_{AA'}\wedge D\pi^A,\label{Edef}
\end{subalign}\noindent
which are the integrals of the Nester\,--\,Witten two-form over a two-dimensional spatial surface $S$ at some finite distance from the source. The spinors $(\bar{\pi}_{A'},\omega^A)$, are once again subject to a specific differential equation, which we will introduce in a minute. In the limit, where the two-surface $S$  reaches spacelike infinity (assuming $S$ have the topology of a two-sphere), the charge $E_\pi$ returns the \textsc{ADM} energy (times $4\pi G$), whilst $Q_{\pi,\omega}$ represents the selfdual part of the relativistic angular momentum plus an additional contribution from the \textsc{ADM} linear momentum, as shown in \eref{QADM}\,---\,provided, though, that the falloff and parity conditions of \hyperref[sec3]{section \ref{sec3}} are satisfied and the asymptotic $r\rightarrow\infty$ expansion of the spinors is given by \eref{taylorom} and \eref{taylorpi}.

To derive balance laws for $E_\pi$ and $Q_{\pi,\omega}$, we employ Stokes's theorem and relate the integrals on different two-surfaces $S_0$ and $S_1$. It is then useful to rewrite the system of equations \eref{witteq1} and \eref{witteq2} in a more covariant form. Indeed, they are completely equivalent to the following set of equations
\begin{subalign}
\frac{1}{2}\tilde{\eta}^{abc}\ou{\Sigma}{A}{Bab}D_c\pi^B=0,\label{witteq1v}\\
\frac{1}{2}\tilde{\eta}^{abc}\ou{\Sigma}{A}{Bab}\left(D_c\omega^B+\ou{\sigma}{B}{B'c}\bar{\pi}^{B'}\right)=0,\label{witteq2v}
\end{subalign}\noindent
where $\ou{\sigma}{AA'}{a}$ are the four-dimensional soldering forms, $\ou{\Sigma}{A}{Bab}$ is the selfdual part \eref{fourdimPauli} of the Pleba\'{n}ski two-form $\Sigma_{\alpha\beta}=e_\alpha\wedge e_\beta$ and $\tilde{\eta}^{abc}$ denotes the \emph{metric independent} Levi-Civita density. The point is then that the equations \eref{witteq1v} and \eref{witteq2v} can be studied on any three-surface, it does not matter whether the three-surface is spacelike, timelike or null, equations \eref{witteq1v} and \eref{witteq2v} always assume the same form. Suppose then the hypersurface is null. On a null-hypersurface $\mathcal{H}$ with topology $\mathcal{H}=[0,1]\times S$ the equations (\ref{witteq1v}, \ref{witteq2v}) turn into
\begin{subalign}
\ell^aD_a\pi^A&=\ell^A\ell_B m^aD_a\pi^B,\label{evolv1}\\
\ell^aD_a\omega^A&=\ell^A\ell_Bm^aD_a\omega^B-3\ell^A\bar{\ell}_{A'}\bar{\pi}^{A'}=0\label{evolv2},
\end{subalign}\noindent
where $\ell^a= \I/2\,\uo{\sigma}{AA'}{a}\ell^A\bar{\ell}^{A'}$ is the null normal in $\mathcal{H}$, which is unique up to overall rescalings, and $m^a$ is the complex null vector, which is defined on a spatial $u=\mathrm{const}.$ section $S_u=\{u\}\times S$ of $\mathcal{H}$ implicitly through: $\sigma_{AA'b}\ell^A\ou{q}{b}{a}=\bar{\ell}_{A'}\bar{m}_b$, where $q_{ab}$ is the pullback of the spacetime metric $g_{ab}$ to the spatial section. The triple $(\ell^a,m^a,\bar{m}^a)$ is a basis in the tangent space $T\mathcal{H}$, and the $SL(2,\C)$  gauge covariant derivative $D_a$ is the pullback of the four-dimensional derivative $\nabla_a$ to the null surface. 

Consider now a four-dimensional spacetime region $\mathcal{B}$, whose boundary is $\partial\mathcal{B}=\mathcal{H}\cup\mathcal{R}_0\cup\mathcal{R}_1$, where $\mathcal{R}_0$ and $\mathcal{R}_1$ are compact spacelike three-surfaces bounded by the two-dimensional \emph{corners} $S_0=\mathcal{H}\cap\mathcal{R}_0$ and $S_1=\mathcal{H}\cap\mathcal{R}_1$. For definiteness, let us assume that $\mathcal{B}$ has the topology of a solid cylinder and that $\mathcal{H}$ is an expanding null surface in the boundary of the causal future of $\mathcal{R}_0$ with topology $\mathcal{H}=[0,1]\times S^2$. We can then foliate $\mathcal{H}$ into $u=\mathrm{const}.$ surfaces $S_u=\{u\}\times S^2$, which have the topology of a two-sphere. The null vector $\ell^a$ is normalised to  $\ell^a\nabla_a u=1$ and future pointing.  The situation is summarised in figure \ref{fig1} below.
 If we were in Minkowski space, $\mathcal{H}$ would be an expanding null surface, which shines out of $S_0$ and hits $S_1$ at some later instance. 
 
 A similar setup has been studied recently by Freidel and Donally, where $\mathcal{B}$ has the shape of a lens, bounded by two spatial hypersurfaces that intersect at a two-dimensional corner \cite{Donnelly:2016auv}.

\begin{figure}[h]
\begin{center}
\includegraphics[width=0.34\textwidth]{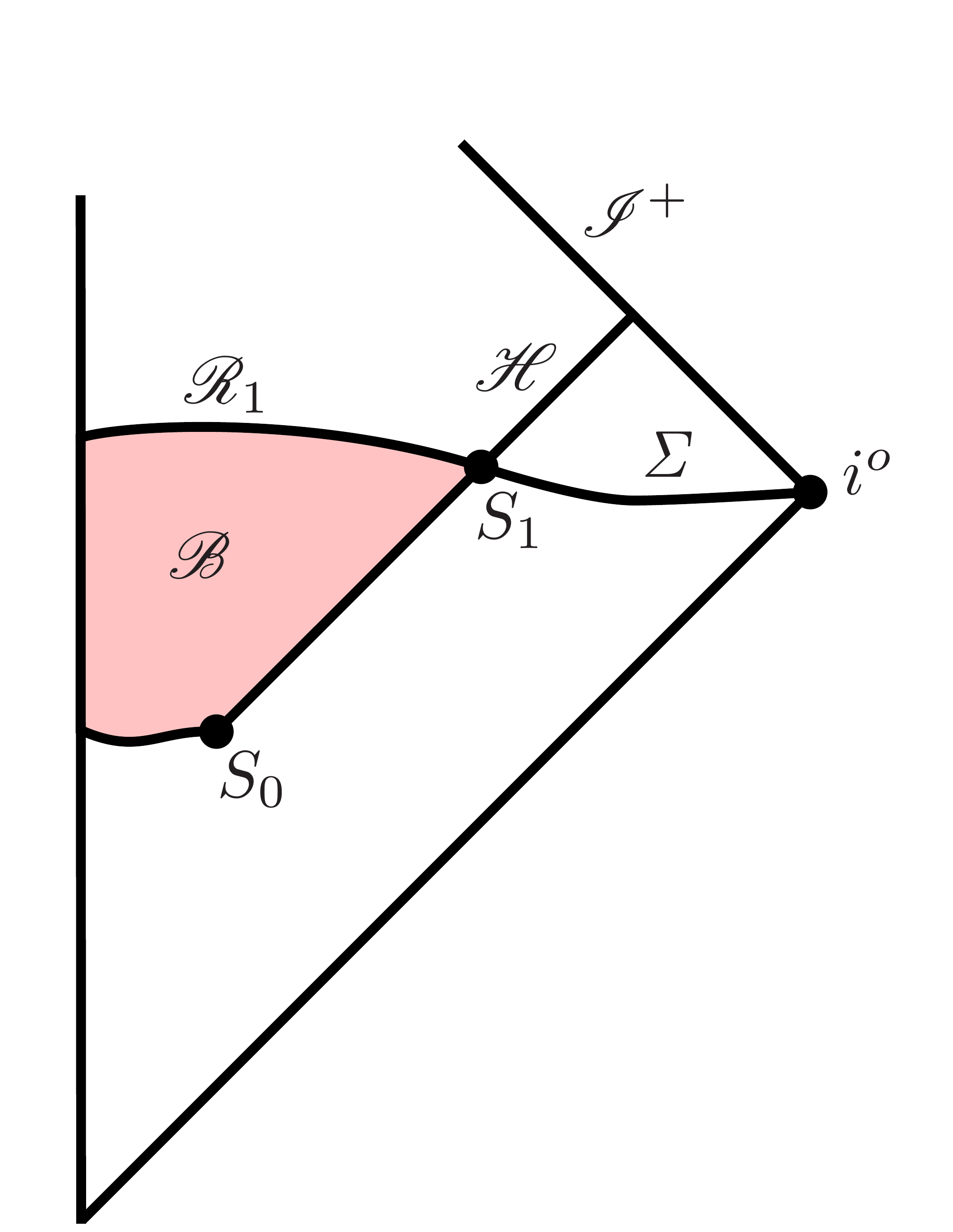}
\end{center}
\caption{The quasi-local charges $E_\pi[S]$ and $Q_{\pi,\omega}[S]$ are associated to the region $\mathcal{B}$ (the system observed), whose boundary contains the expanding null surface $\mathcal{H}$. The charges $E_\pi[S]$ and $Q_{\pi,\omega}[S]$  are surface integrals over  two-dimensional cross-sections of $\mathcal{H}$. The spinors $(\bar{\pi}_{A'},\omega^A)$ are assumed to solve the generalised Witten equations (\ref{witteq1v}, \ref{witteq2v}) along the entire boundary $\partial\mathcal{B}$. At finite distance, infinitely many such charges are expected. In the limit where $S$ goes to spacelike infinity $i_o$ only ten of them survive: The linear momentum, the angular momentum and the centre of mass.}\label{fig1}
\end{figure}

We will also assume that the equations \eref{witteq1v} and \eref{witteq2v} admit a solution along the \emph{entire} boundary of $\mathcal{B}$. It goes well beyond this article to study under which conditions this is possible, but the idea is the following: We can certainly find a solution $(\bar{\pi}_{A'},\omega^A)_{o}$ of (\ref{witteq1v}, \ref{witteq2v}) on the spacelike bottom $\mathcal{R}_0$ of $\partial\mathcal{B}$. We then use the evolution equations (\ref{evolv1}, \ref{evolv2}) to propagate this solution along the null generators of the surface $\mathcal{H}$. We have thus found a solution in $\mathcal{R}_0\cup\mathcal{H}\subset\partial\mathcal{B}$. The question is then how to extend this solution to the top $\mathcal{R}_1$ of our cylinder. This may be achieved by using the ADM Hamiltonian and propagate the solution from the boundary into the bulk. We would thus foliate the interior of $\mathcal{B}$ by a family of spatial hypersurfaces $\mathcal{D}_t$ that all intersect at the top corner $S_1=\mathcal{H}\cap\mathcal{R}_1$, such that $\lim_{t\rightarrow 0}\mathcal{D}_t=\varSigma_0\cup\mathcal{H}$ and $\lim_{t\rightarrow 1}\mathcal{D}_t=\mathcal{R}_1$. A given solution of the generalised Witten equations \eref{witteq1v} and \eref{witteq2v} on some initial hypersurface $\mathcal{D}_{t_o}$ of $\mathcal{B}$ is then determined by two kinds of initial data: By the initial value $(\vu{\bar\pi}{0}{A'},\vo{\omega}{0}{A})$ of the spinors at the corner $S_1$, and the value of the boundary fields\footnote{Notice, that the Ashtekar variables $\ou{A}{A}{Ba}$ and $\ou{E}{AB}{a}=\frac{1}{2}\tilde{\eta}^{abc}\ou{\Sigma}{AB}{bc}$, which appear in the generalised Witten equations (\ref{witteq1v}, \ref{witteq2v}) are canonically conjugate variables  on the slices of the foliation: $\{\uo{E}{AB}{a}(p),\ou{A}{CD}{b}(q)\}=-4\pi\I G\delta^a_b\delta^{(C}_A\delta^{D)}_B\delta^{(3)}(p,q)$. Using Ashtekar's Hamiltonian \cite{ashtekar,Newvariables}, we can then propagate solutions of the generalised Witten equations from the boundary into the bulk $\mathcal{B}$.} 
$\ou{A}{A}{Ba}$ and $\frac{1}{2}\tilde{\eta}^{abc}\ou{\Sigma}{AB}{bc}$ on $\mathcal{D}_{t_o}$. Any such solution of the Witten equations on $\mathcal{D}_{t_0}$ can be thus written as a functional
\begin{equation}
\pi^A\big[A,E,\vu{\bar\pi}{0}{}\big],\quad\mbox{resp.}\quad\omega^A\big[A,E,\vu{\bar\pi}{0}{},\vo{\omega}{0}{}\big]
\end{equation}
of the boundary data $\ou{A}{A}{Ba}$ and $\uo{E}{A}{Ba}$ on $\mathcal{D}_{t_0}$ and the boundary data $(\vu{\bar\pi}{0}{A'},\vo{\omega}{0}{A})$ on $S_1=\partial\mathcal{D}_{t_o}$. We can then use the gravitational Hamiltonian (for a given choice of lapse $N$ and shift $N^a$ that vanish at the corner: $N=0$ and $N^a=0$ at $S_1=\partial\mathcal{D}_t$) to evolve this solution into the bulk, thus obtaining a $t$-parameter family of spinors $(\bar{\pi}_{A'},\omega^A)_t$, which functionally depend on the initial values of both the spinors $(\vu{\bar\pi}{0}{A'},\vo{\omega}{0}{A})$ at $S_1=\partial\mathcal{D}_{t_o}$ and the canonical variables $\ou{A}{A}{Ba}$ and $\uo{E}{A}{Ba}$ on $\mathcal{D}_{t_o}$. If $(\bar{\pi}_{A'},\omega^A)_t$ is then a solution of the generalised Witten equations (\ref{witteq1v}, \ref{witteq2v}) on the initial slice $\mathcal{D}_{t_o}$, it will also be a solution on some later slice $\mathcal{D}_{t_o+\varepsilon}$. Taking the limits ${t\rightarrow 0,1}$ we can the  extend the solution from $\mathcal{R}_0\cup\mathcal{H}$ all around the boundary of $\mathcal{B}$. The only difficulty arises at the corner $S_1$, where the Hamiltonian vanishes only up to a local $SL(2,\C)$ gauge transformation, which can however always be set to zero. At the moment, this is only a very formal argument, a more careful Hamiltonian analysis will be presented in a forthcoming paper.


If we then have such a solution of the evolution equations \eref{witteq1v} and \eref{witteq2v} along $\mathcal{H}$, we can immediately write the difference between the charges at two consecutive moments of $u$ as a volume integral over the null surface $\mathcal{H}$ between, indeed
\begin{eqnarray}\nonumber\fl\qquad
&Q_{\pi,\omega}[S_1]-Q_{\pi,\omega}[S_0]=\I\int_{\mathcal{H}}\left[\bar{\omega}^{A'}\sigma_{AA'}\wedge\ou{F}{A}{B}\pi^B-D\bar{\omega}^{A'}\wedge\sigma_{AA'}\wedge D\pi^A\right],
\end{eqnarray}
which is a consequence of Stokes's theorem and the definition of the curvature two-form as the square of the exterior covariant derivative. We simplify this equation and insert the Einstein equations for a given energy-momentum tensor $T_{ab}$ together with the evolution equations \eref{witteq1v} and \eref{witteq2v} for the spinors $(\bar{\pi}_{A'},\omega^A)$ along the null surface $\mathcal{H}$. After some straight forward manipulations, we then find the following conservation law for the quasi-local angular momentum
\begin{eqnarray}\nonumber\fl\qquad
&Q_{\pi,\omega}[S_1]-Q_{\pi,\omega}[S_0]=-4\pi G\int_{\mathcal{H}}\di u\wedge d^2v\,T_{AA'BB'}\ell^A\bar{\ell}^{A'}\pi^B\bar{\omega}^{B'}+\\
\fl&\qquad\quad+\int_{\mathcal{H}}\di u\wedge d^2v\left[q^{ab}D_a\bar{\omega}^{A'}D_b\pi^A\bar{\ell}_{A'}\ell_A+\ell^a\ell^bD_a\bar\omega^{A'}D_b\pi^A \bar{k}_{A'}k_A\right].\label{angularflux}
\end{eqnarray}
The derivation follows exactly the same algebraic steps that brought us to \eref{Qvol2}, the only difficulty is that the hypersurfcace $\mathcal{H}$ is null rather than spacelike. Equation \eref{angularflux} has an immediate physical interpretation. The first term represents the contribution to the asymptotic charge \eref{QADM} from matter falling into $\mathcal{B}$, the second term represents the influx of gravitational radiation, but contributions from the purely geometrical expansion of $\mathcal{H}$ in $M$ are also possible\,---\,if, for instance, $\mathcal{H}$ is an expanding light cone in Minkowski space, one can always construct solutions of (\ref{evolv1}, \ref{evolv2}), for which $Q_{\pi,\omega}$ is not constant. The notation should be clear: $T_{AA'BB'}=-\frac{1}{2}T_{ab}\uo{\sigma}{AA'}{a}\uo{\sigma}{BB'}{b}$ is the spinorial version of the stress-energy tensor $T_{ab}$, the dyad $(\ell^A,k^A)$ is normalised to $k_A\ell^A=1$, $q_{ab}=m_{(a}\bar{m}_{b)}$ is the two-metric on the $u=\mathrm{const}.$ cross-sections $S_u$ of $\mathcal{H}$ and $d^2v=\frac{1}{2\I}\tilde{\eta}^{ab}m_a\bar{m}_b$ is the corresponding volume element.

The balance law for the quasi-local energy can be derived in a similar manner, but let us first note  that $E_\pi[S_0]$ is always real, which follows from its definition \eref{Edef} by partial integration. Indeed
\begin{eqnarray}\nonumber
\bar{E}_\pi[S]&=\I\int_S\pi^A\sigma_{AA'}\wedge D\bar{\pi}^{A'}=+\I\int_SD\pi^A\wedge\sigma_{AA'}\bar{\pi}^{A'}\\
&=-\I\int_S\bar{\pi}^{A'}\wedge\sigma_{AA'}D\pi^A=E_\pi[S].
\end{eqnarray}
To derive a balance law, we then employ Stokes's theorem and write the difference $E_{\pi}[S_1]-E_{\pi}[S_0]$ as a volume integral over the null surface $\mathcal{H}$. We insert the Einstein equations together with the generalised Witten equations (\ref{evolv1}, \ref{evolv2}) for the propagation of $(\bar{\pi}_{A'},\omega^A)$ along $\mathcal{H}$, which eventually leads us to
\begin{eqnarray}\nonumber\fl\quad&
E_{\pi}[S_1]-E_{\pi}[S_0]=
-4\pi G\int_{\mathcal{H}}\di u\wedge d^2v\,T_{AA'BB'}\ell^A\bar{\ell}^{A'}\pi^B\bar{\pi}^{B'}+\\
\fl&\qquad\quad\,+\int_{\mathcal{H}}\di u\wedge d^2v\left[q^{ab}D_a\bar{\pi}^{A'}D_b\pi^A\bar{\ell}_{A'}\ell_A+\ell^a\ell^bD_a\bar\pi^{A'}D_b\pi^A \bar{k}_{A'}k_A\right].\label{Ebalance}
\end{eqnarray}
The first term represents the influx of matter, the second term describes the contribution to the \textsc{ADM} energy \eref{ADMen} from gravitational radiation falling into $\mathcal{B}$, but contributions from the purely geometrical expansion of $\mathcal{H}$ in $M$ are also possible (we will come back to this point below). 
The first term is positive provided the four-vector $-\ou{T}{a}{b}\ell^b\big|_{\mathcal{H}}$ is causal (i.e.\ future directed, and time-like or null), which we will always assume in the following, the second term is always positive (or respectively zero). The quasi-local energy $E_\pi[S]$ can therefore only increase in time
\begin{equation}
E_{\pi}[S_1]\geq E_{\pi}[S_0]\geq 0,
\end{equation}
where the second inequality follows from Witten's original argument \cite{Wittenproof}, see also \eref{postvty}. Similar equations have been derived by Ludvigsen and Vilckers \cite{LudvigsenAnglrMntm} using, however, a propagation law for the spinors, which is different from \eref{evolv1}, but this is unnecessary: One does not need to specify an additional propagation law for the spinors on top of the generalised Witten equations \eref{witteq1v} and \eref{witteq2v}. The generalised Witten equations are strong enough to extend the spinors from a given two-dimensional cross-section $S_u\subset\partial\mathcal{B}$ along the entire three-boundary $\partial\mathcal{B}\supset S_u$ (provided a solution exists). Whether $\partial\mathcal{B}$ is spacelike, timelike or null does not matter for the construction. 

\paragraph{Under which conditions are the quasi-local charges conserved?} Suppose now that for a given solution $(\bar{\pi}_{A'},\omega^A)$ of (\ref{evolv1}, \ref{evolv2}) the quasi-local energy $E_\pi[S_u]$ is constant in $u$. Under which conditions is this possible? First of all, we demand that the four-vector $j^a=-\ou{T}{a}{b}\ell^b\big|_{\mathcal{H}}$, is causal (i.e.\ future directed, and time-like or null), in which case all terms on the right hand side of the balance law \eref{Ebalance} are greater or equal to zero. When do they then vanish? For the first term to vanish the energy flux $j^a=-\ou{T}{a}{b}\ell^b$ must be proportional to the null vector $\sigma_{AA'a}\pi^A\bar{\pi}^{A'}$ itself, hence
\begin{equation}
	\forall u:\frac{\di}{\di u}E_\pi[S_u]=0\Rightarrow T_{AA'BB'}\ell^B\bar{\ell}^{B'}\big|_{\mathcal{H}}\propto\pi_{A}\bar{\pi}_{A'}.\label{cons1}
\end{equation}
For the second term to vanish $\uo{q}{a}{b}D_b\pi^A$ must be of the form $\omega_a\ell^A$, where $\omega_a$ is a complex-valued one-form on $\mathcal{H}$. But we also know that $\pi^A$ must solve the evolution equations \eref{evolv1}. The condition $\uo{q}{a}{b}D_b\pi^A=\omega_a\ell^A$ is compatible with the evolution equations on $\mathcal{H}$ only if $\ell^aD_a\pi^A=0$, in which case the third term in the balance \eref{Ebalance} vanishes as well. We have thus shown that
\begin{equation}
	\forall u:\frac{\di}{\di u}E_\pi[S_u]=0\Rightarrow\exists\omega_a\in T^\ast\mathcal{H}_\C:D_a\pi^A=\omega_a\ell^A,\;\ell^a\omega_a=0.\label{cons2}
\end{equation}
Going back to the equation \eref{angularflux} for the influx of angular momentum into $\mathcal{B}$, we then see that the equations \eref{cons1} and \eref{cons2} also imply that the quasi-local angular momentum \eref{Qdefvar} is preserved as well, hence
\begin{equation}
\forall u:\frac{\di}{\di u}E_\pi[S_u]=0\Rightarrow\forall u:\frac{\di}{\di u}Q_{\pi,\omega}[S_u]=0.\label{cons3}
\end{equation}

	So far, we have asked under which conditions the quasi-local energy $E_\pi[S_u]$ is conserved for \emph{one} particular solution $(\bar{\pi}_{A'},\omega^A)$ of the generalised Witten equations (\ref{witteq1v}, \ref{witteq2v}). Consider now two linearly independent solutions $(\bar{o}_{A'},\omega^A_o)$ and $(\bar{\iota}_{A'},\omega^A_\iota)$ of \eref{evolv1} such that $\iota_Ao^A\neq 0$ on $\mathcal{H}$. 
What are then the conditions under which the quasi-local energy $E_\pi[S_u]$ for both $o^A$ and $\iota^A$ is constant along the null generators of $\mathcal{H}$? First of all, we return to the balance law \eref{Ebalance} for the quasi-local energy. The spinors $o^A$ and $\iota^A$ are assumed to be linearly independent,  which implies that the first term in \eref{Ebalance} vanishes for both of them only if there is no influx of matter, hence
	\begin{equation}
T_{ab}\ell^b\big|_{\mathcal{H}}=0,\label{nomatterflux}
\end{equation}
which is a consequence of our requirement that the energy flux $j^a=-\ou{T}{a}{b}\ell^b|_{\mathcal{H}}$ is causal  (i.e.\ future directed, and time-like or null) on $\mathcal{H}$.	
	
For the second term in \eref{Ebalance}, we repeat the argument that brought us to \eref{cons2}. We then see that $\uo{q}{a}{b}D_b\pi_A\ell^A$ can vanish for both $\pi^A=\iota^A$ and $\pi^A=o^A$ only if $\uo{q}{a}{b}D_bo^A\propto\ell^A$ and $\uo{q}{a}{b}D_b\iota^A\propto\ell^A$. This implies that there must exist complex-valued one-forms $\omega_a[\iota]$ and $\omega_a[o]$ in $T^\ast\mathcal{H}_\C$ such that
\begin{subalign}
D_a o^A=\omega_a[o]\ell^A,&\quad\ell^a\omega_a[o]=0,\quad\mbox{and}\label{spinrot1}\\
D_a \iota^A=\omega_a[\iota]\ell^A,&\quad\ell^a\omega_a[\iota]=0.\label{spinrot2}
\end{subalign}\noindent
That $\ell^a\omega_a[\cdot]=0$ is a consequence of our initial assumption that $o^a$ and $\iota^A$ are solutions of the evolution equation \eref{evolv1} along $\mathcal{H}$.
Notice, that these equations immediately imply that the product $\iota_A o^A$ is preserved along the null generators, which is consistent with our initial assumption: The spinors $(o^A,\iota^A)$ are a linearly independent dyad all along the entire null surface $\mathcal{H}$.

Finally, we can use \eref{spinrot1} and \eref{spinrot2} to compute the components of the curvature tensor with respect to the coordinates $(u,z,\bar{z})$ on $\mathcal{H}$, which are constructed as follows: $\ell^a=\partial^a_u$ is the null vector in $\mathcal{H}$ and $u$ is an affine parameter along the null rays of $\mathcal{H}$, hence $\partial^b_uD_b\partial^a_u=0$. The tangent vectors $\partial_z^a$ and $\partial_{\bar{z}}^a$ in $ [TS_u]_{\C}$, on the other hand, are complex null vectors on the transversal $u=\mathrm{const}.$ slices $S_u$ of $\mathcal{H}$. 

From $\ell^bD_b\ell^a=0$ and $\ell^a\propto \uo{\sigma}{AA'}{a}\ell^A\bar{\ell}^{A'}$ we then also know that the spinor $\ell^A$ has a covariant derivative $\ell^aD_a\ell^A\propto\ell^A$. Going back to the equations \eref{spinrot1} and \eref{spinrot2} for the spin-rotation coefficients, we thus get
\begin{subalign}
\big[D_{\partial_u},D_{\partial_z}\big]o^A=\ell^aD_{a}\big(D_{\partial_z}o^A\big)=\ell^aD_{a}\big(\omega_b[o]\partial^b_z \ell^A\big)\propto \ell^A,\\
\big[D_{\partial_u},D_{\partial_z}\big]\iota^A=\ell^aD_{a}\big(D_{\partial_z}\iota^A\big)=\ell^aD_{a}\big(\omega_b[\iota]\partial^b_z \ell^A\big)\propto \ell^A.
\end{subalign}\noindent
This system of equations constrains the $(u,z)$ and $(u,\bar{z})$ components of the selfdual curvature two-form $F_{ABab}$ on $\mathcal{H}$. Indeed we get
\begin{equation}
F_{ABab}\ell^a\partial^b_z\propto\ell_{(A}\ell_{B)},\quad\mbox{and}\quad
F_{ABab}\ell^a\partial^b_{\bar{z}}\propto\ell_{(A}\ell_{B)}.\label{Fcomps}
\end{equation}
With no matter falling into $\mathcal{B}$, as implied by \eref{nomatterflux}, this implies for the corresponding components of the Weyl tensor that
 \begin{equation}
C_{abcd}\ell^b\ell^c\partial^c_z=0,\quad C_{abcd}\ell^b\ell^c\partial^c_{\bar z}=0,
\end{equation}
which is the same as to say that an observer that follows the null generators of $\mathcal{H}$ does not experience any tidal forces, and does, therefore, not experience any influx of gravitational radiation. Going back to the balance law \eref{angularflux} for the inflow of angular momentum, and employing equations \eref{spinrot1} and \eref{spinrot2} for the covariant derivatives of $o^A$ and $\iota^A$, we then see that the quasi-local angular momentum is conserved  as well. We have thus convinced ourselves of the following: 
\vspace{0.2em}

\emph{There is no influx of matter or gravitational radiation across  the expanding null surface $\mathcal{H}$, if there exist two linearly independent solutions $o^A$ and $\iota^A$ (i.e.\ $\iota_A o^A\neq 0$ on $\mathcal{H}$) of the first evolution equation \eref{witteq1v}, such that the $\iota^A$ and $o^A$ components $E_\iota[S_u]$ and $E_o[S_u]$ of the quasi-local energy $E_\pi[S_u]$ are constant along the null generators of $\mathcal{H}$. In addition, the quasi-local angular momentum $Q_{\pi,\omega}$ for corresponding solutions $(\bar{o}_{A'},\omega^A_o)$ and $(\bar{\iota}_{A'},\omega^A_\iota)$ of the second evolution equation \eref{witteq2v} is preserved as well.}
\vspace{0.2em}

The converse is however false, for even in Minkowski space, one can always find some spinors $(\bar{\pi}_{A'},\omega^A)$ that solve both \eref{evolv1} and \eref{evolv2} on $\mathcal{H}$ whose quasi-local charge $E_\pi[S_u]$ grows over all limits with $u\rightarrow\infty$, which simply reflects the purely geometric growth in area of the expanding cross-sections $S_u$ of $\mathcal{H}$ in $M$.

\paragraph{} We have called $E_\pi[S]$ and $Q_{\pi,\omega}[S]$ quasi-local charges for energy and angular momentum. This terminology is clearly borrowed from our intuition at spacelike infinity, but we have to be careful, there is a crucial difference. At spacelike infinity, there are only ten conserved charges, corresponding to boosts, rotations and translations. At a finite distance, the situation is very different. Now, we have infinitely many quasi-local charges $E_\pi[S]$ and $Q_{\pi,\omega}[S]$. The reason is the following: The system of first-order partial differential equations \eref{witteq1v} and \eref{witteq2v} on $\partial\mathcal{B}=\mathcal{H}\cup\mathcal{R}_0\cup\mathcal{R}_1$ is characterised by an initial datum on some $u=\mathrm{const}.$ section $S_u$ of $\mathcal{H}$. The state space of all initial data $(\vu{\bar\pi}{0}{A'},\vo{\omega}{0}{A})$ at $S_u$, which are admissible with \eref{witteq1v} and \eref{witteq2v} is most likely infinite dimensional, and there would be then infinitely many corresponding charges $E_\pi[S]$ and $Q_{\pi,\omega}[S]$. Following Dougan an Mason \cite{MasonEnergy} we could choose to reduce the functional dependence of the spinors by demanding that $\pi^A$ is e.g.\ holomorphic, i.e.\ $m^aD_a\pi^A=0$ on the $u=0$ initial hypersurface $S_0\subset\mathcal{H}$. This would dramatically reduce the functional freedom (depending on the topology of the cross-sections $S_u$ of $\mathcal{H}$) down to  (generically) two linearly independent solutions for $\pi^A$. This is however a rather ad hoc restriction, physically unjustified: In general, the evolution equations \eref{evolv1} will not preserve the holomorphicity of $\pi^A$. For if $\pi^A$ were holomorphic on $S_u$, it would not be holomorphic on $S_{u+\varepsilon}$, unless $\ou{F}{A}{Bab}\ell^am^b=0$, which is unphysical, because it excludes gravitational radiation falling into $\mathcal{B}$. We should therefore allow all possible initial data on $S_u$, which are presumably infinitely many. In fact, it should not come as a surprise. General relativity is background independent, and the relevant local gauge group is the entire diffeomorphism group over $\mathcal{B}$. This group is infinite dimensional, hence there should be infinitely many charges associated to it.

That only a finite number of them survive at spacelike infinity is a consequence of the falloff and parity conditions. If the parity conditions are satisfied, we can restrict ourselves to spinors $(\bar{\pi}_{A'},\omega^A)$, whose asymptotic $r\rightarrow\infty$ expansion is governed by \eref{taylorpi} and \eref{taylorom} (where the leading order terms $\vo{\pi}{1}{A}(\vartheta,\varphi)$ and $\varphi^A(\vartheta,\varphi)$ in the asymptotic expansion \eref{taylorpi} and \eref{taylorom} are respectively even and odd), for all other spinors that solve the three-surface spinor equations (\ref{witteq1v}, \ref{witteq2v}), but violate the parity conditions and falloff conditions, the integrals at infinity would either vanish or diverge. The only charges that remain at spacelike infinity are then the linear momentum, the angular momentum and the centre of mass.

\section{Summary, discussion, outlook}\label{sec6}
We have shown in this paper that Witten's construction of the \textsc{ADM} energy can be generalised to the gravitational angular momentum and centre of mass. This is achieved by once iterating Witten's equation, i.e.\ by solving it for a source that is itself a solution of the Witten equation. The resulting system (\ref{witteq1}, \ref{witteq2}) of elliptic partial differential equations was studied around spacelike infinity. We imposed parity conditions\footnote{The leading $O(r^{-1})$ term of the metric perturbation is parity even, and the leading $O(r^{-2})$ order of the difference tensor between the Ashtekar\,--\,Sen connection and the fiducial flat connection at infinity is parity odd.} on metric and connection (which are needed to remove otherwise divergent terms from the definition of the angular momentum, see \cite{Corichi:2015cqa, Ashtekar:1978zz, ReggeTeitelboim} for references) and solved the system (\ref{witteq1}, \ref{witteq2}) of equations with the ansatz \eref{taylorpi} and \eref{taylorom} for the perturbative $r\rightarrow\infty$ expansion of the spinors $(\bar{\pi}_{A'},\omega^A)$. We showed that solutions exist for which $\bar{\pi}_{A'}$ is asymptotically constant, while $\omega^A$ grows linearly with $r$.  Next, we inserted the solutions into the integral \eref{Qcharge2} over the Nester\,--\,Witten two-form and evaluated the charge $Q_{\pi,\omega}$ in the limit of $r\rightarrow\infty$. Several terms vanish thanks to the parity conditions and the only terms remaining organise themselves into a complex linear combination of all ten Poincaré charges \eref{QADM}  at spacelike infinity. 

The result improves on a paper by Shaw \cite{0264-9381-2-2-012}. Shaw was mainly interested in null infinity, and gave a heuristic argument (which is confirmed by this paper) why the solutions to the generalised Witten equations (\ref{witteq1}, \ref{witteq2}) should return both angular momentum and centre of mass at spacelike infinity.

Finally, we saw that the generalised Witten equations (\ref{witteq1}, \ref{witteq2}) are equivalent to the equations (\ref{witteq1v}, \ref{witteq2v}), which require the vanishing of certain spinor-valued three-forms. These equations can be studied, therefore, on a three-dimensional submanifold of arbitrary signature. We argued\footnote{The argument was the following: We can certainly find a solution at either of the two spacelike hypersurfaces bounding $\partial\mathcal{B}$ (see figure \ref{fig1} for an illustration). The evolution equations \eref{evolv1} and \eref{evolv2} propagate this solution uniquely along the null generators of $\mathcal{H}$. One should then employ the Hamiltonian of GR, as written in terms of the selfdual Ashtekar variables \cite{ashtekar,Newvariables}, and extend the solution into the interior of $\mathcal{B}$, which will induce a solution along the entire three-boundary of $\mathcal{B}$.} that the system of equations (\ref{witteq1v}, \ref{witteq2v}) admits solutions along the entire three-boundary $\partial\mathcal{B}$ of a compact four-dimensional region $\mathcal{B}$, which has the topology of a cylinder, whose top and bottom parts are spacelike hypersurfaces, whereas the side is a null surface $\mathcal{H}=[0,1]\times S^2$. Next, we proposed quasi-local charges $Q_{\pi,\omega}[S_u]$ and $E_\pi[S_u]$, which are two-dimensional surface integrals over an arbitrary cross-section of $\mathcal{H}$. Using Stokes's theorem on $\mathcal{H}$ we proved balance laws for $E_\pi[S]$ and $Q_{\pi,\omega}[S]$. If $\mathcal{H}$ is an outgoing null surface, the \emph{infinitely many} quasi-local charges $E_\pi[S_u]$ can only increase along the null generators $\ell^a$ of $\mathcal{H}$ (provided the energy flux $j^a=-\ou{T}{a}{b}\ell^b\big|_{\mathcal{H}}$ is causal). We also saw that the charges capture the influx of matter and radiation: If there are two linearly independent solutions of the evolution equation \eref{witteq1v}, for each of which the quasi-local energy $E_\pi[S_u]$ is preserved along the null generators of $
\mathcal{H}$, then there will be no influx of matter or gravitational radiation into the bulk. The corresponding quasi-local angular momentum $Q_{\pi,\omega}[S]$ will be conserved as well. 

At finite distance, there are infinitely many such charges on a given two-surface $S$, because there is no preferred choice of spinors $\bar{\pi}_{A'}$ and $\omega^A$ along the entire null surface shining out of $S$. At spacelike infinity only ten of them survive. All other quasi-local charges either diverge or vanish in the limit of $S$ reaching spacelike infinity. At null infinity, the situation is however very different. There is no unambiguous definition of angular momentum, and we can expect that infinitely many such quasi-local charges $Q_{\pi,\omega}[S_u]$ and $E_{\pi}[S_u]$ survive the limit towards null infinity. 

This observation and the relevance of super-translations for the appearance of a parity odd $O(r^0)$ term in the asymptotic  expansion \eref{taylorom} of the $\omega^A$-spinor suggests a relation between the quasi-local charges defined by the generalised Witten spinors $\bar{\pi}_{A'}$ and $\omega^A$ and the canonical charges for the BMS group \cite{Ashtekar:1990gc}. At the moment, the precise connection has not been worked out, but the basic idea that could establish such a correspondence is the following. In the covariant Hamiltonian formalism,  quasi-conserved quantities \cite{Wald:1999wa} are defined by the Noether current three-form, which can be inferred from the boundary plus bulk action. Suppose then that the boundary has a component that is null (such as in figure \ref{fig1} above). On a null surface, working with self-dual variables \cite{Newvariables}, one may use a gravitational boundary term with spinors as the fundamental boundary variables, c.f.\ \cite{Wieland:2016aa}. The canonical Noether charge (as derived from the boundary plus bulk action) would be then a two-dimensional surface integral over certain boundary spinors and their derivatives. If one would then consider a sequence of null boundaries approaching null infinity, it seems likely that the resulting Noether charges for the asymptotic BMS symmetries agree with the generalised Nester\,--\,Witten charges defined in this paper. Clearly, the details must be worked out more carefully, and we leave a more complete discussion for future research.

\paragraph{} The paper may be relevant for quantum gravity as well. 
In quantum gravity, the quasi-local charges for quasi-local energy and angular momentum will turn into operators $\widehat{E}_{\pi}[S]$ and $\widehat{Q}_{\pi,\omega}[S]$ on a Hilbert space.
For any choice of spinors $(\bar{\pi}_{A'}(\vartheta,\varphi),\omega^A(\vartheta,\varphi))$ on $S$, we can then imagine the notion of a $(\bar{\pi}_{A'},\omega^A)$-vacuum as a coherent state in the kernel of the corresponding quasi-local energy
\begin{equation}
\widehat{E}_{\pi}[S]\bigl|\bar{\pi}_{A'}(\vartheta,\varphi),\omega^A(\vartheta,\varphi)\bigr\rangle=0.\label{vacuumstate}
\end{equation}
We expect that there are infinitely many such states, because there are infinitely many choices for the spinors $(\bar{\pi}_{A'}(\vartheta,\varphi),\omega^A(\vartheta,\varphi))$ on $S$, and this degeneracy resonates with recent ideas by Hawking, Perry and Strominger \cite{Hawking:2016msc} in the context of black hole thermodynamics. 

The evolution of such a boundary state \eref{vacuumstate} along the null surface shining out of $S$ will be determined by the quantisation of the balance laws \eref{angularflux} and \eref{Ebalance} for both $Q_{\pi,\omega}[S]$ and $E_\pi[S]$. 
These balance laws are, in fact, nothing but the pull-back of the Einstein equations integrated over the null surface. The quantisation of this infinite tower of constraint equations defines, therefore, a version of the Wheeler\,--\,De\,Witt equation, but on a null surface $\mathcal{H}$ rather than a spacelike hypersurface. The boundary spinors $(\bar\pi_{A'},\omega^A)$ that crucially enter the classical balance laws as smearing functions  should be then quantised as well. This idea is supported by a very recent paper \cite{Wieland:2016aa}, where it is shown that the natural boundary variables for the gravitational field on a null surface are a spinor and its canonical momentum conjugate, which is, in four space-time dimensions, a spinor-valued two-form. This idea also resonates with recent developments in loop quantum gravity. During the last couple of years the Hilbert space built over the Ashtekar\,--\,Lewandowski vacuum \cite{LOSTtheorem} was recast in terms of spinors, which treat both the group-valued holonomies and the Lie algebra-valued fluxes in a symmetric way \cite{twist,Dupuis:2012vp,komplexspinors,twistintegrals,Bianchi:2016hmk,Borja:2010rc}.

\appendix
\renewcommand{\theequation}{A.\arabic{equation}}
\section*{Acknowledgements}
I thank Abhay Ashtekar and Eugenio Bianchi for many valuable comments and the hospitality at the Institute for Gravitation and the Cosmos. 
I thank my colleagues from the Perimeter Institute, in particular Aldo Riello, Laurent Freidel, Henrique Gomes and Lee Smolin for many enlightening discussions, helpful comments and critical remarks. 
\emph{Research at Perimeter Institute is supported by the Government of Canada through Industry Canada and by the Province of Canada through the Ministry of Research and Innovation.}
\section*{Appendix}\label{appdx}
$A,B,C,\dots$ are spinor indices, they are raised and lowered by the two-dimensional epsilon tensors $\epsilon_{AB}$ and $\epsilon^{CD}$ according to: $\pi_A=\pi^B\epsilon_{BA}$ and $\pi^A=\epsilon^{AB}\pi_B$. Each spinor carries the fundamental spin $(\frac{1}{2},0)$ representation of the universal cover of the Lorentz group: $g\triangleright\pi^A=\ou{g}{A}{B}\pi^B$ for any $g\in SL(2,\C)$. Primed indices $A',B',C',\dots$ refer to the complex conjugate spin $(0,\frac{1}{2})$ representation $g\triangleright\bar\pi^{A'}=\ou{\bar g}{A'}{B'}\bar\pi^{B'}=\overline{\ou{g}{A}{B}\pi^B}$.

Internal Lorentz indices are denoted with Greek indices $\alpha,\beta,\dots$\,, the signature of the Minkowski metric $\eta_{\alpha\beta}$ is ($-$$+$$+$$+$). The Levi-Civita tensor is $\epsilon_{\alpha\beta\mu\nu}$ with $\epsilon_{\mu\nu\alpha\beta}T^\mu X^\nu Y^\alpha Z^\beta>0$ for any positively oriented quadruple of four-vectors $(T^\alpha,X^\alpha,Y^\alpha,Z^\alpha)$.  

$a,b,c,\dots$ are abstract tensor indices for either all of spacetime $M$, an initial  surface $\varSigma$ or spacelike infinity.  Conversely, $\ou{e}{\alpha}{a}$ denotes a tetrad, and its pullback to $\varSigma$ defines a triad $\ou{e}{i}{a}$, where $i,j,k,\dots$ are internal $\mathfrak{su}(2)_n$ indices with respect to the surface normal $n^\alpha=\ou{e}{\alpha}{a}n^a$ of $\varSigma$. The metric-independent Levi-Civita tensor densities are $\tilde{\eta}^{abcd}$, $\tilde{\eta}^{abc}$ and $\tilde{\eta}^{ab}$ in four, three and two dimensions respectively.

Finally, we have the soldering forms $\ou{\sigma}{AA'}{\alpha}$. An explicit representation is given by the quadruple of Hermitian matrices
\begin{equation}
\fl\quad\left[\ou{\sigma}{AA'}{0},\ou{\sigma}{AA'}{1},\ou{\sigma}{AA'}{2},\ou{\sigma}{AA'}{3}\right]=
\left[\biggl(\begin{array}{rr} 
1&\phantom{+}0\\0&1
\end{array}\biggr),
\biggl(\begin{array}{rr} 
0&\phantom{+}1\\1&0
\end{array}\biggr),
\biggl(\begin{array}{rr} 
0&-\I\\\I&0
\end{array}\biggr),
\biggl(\begin{array}{rr} 
1&0\\0&-1
\end{array}\biggr)\right],
\end{equation}
where $A$ and $A'$ refer to the row and column indices, which transform under the spin $(\frac{1}{2},0)$ and $(0,\frac{1}{2})$ representations of $SL(2,\C)$. The soldering forms $\ou{\sigma}{AA'}{\alpha}$ satisfy the following important identity, which is the generalisation of the Pauli identity $\sigma_i\sigma_j=\delta_{ij}\mathds{1}+\I\uo{\epsilon}{ij}{k}\sigma_k$ to four dimensions,
\begin{equation}\fl\quad
\ou{\sigma}{AC'}{\alpha}\bar{\sigma}_{C' B\beta}=-\delta^A_B\,\eta_{\alpha\beta}-2\ou{\Sigma}{A}{B\alpha\beta},\quad
\ou{\Sigma}{AB}{\alpha\beta}:=\frac{1}{2}\ou{\sigma}{A}{C'[\alpha}\ou{\bar\sigma}{C'B}{\beta]},\label{fourdimPauli}
\end{equation}
where $\ou{\Sigma}{A}{B\alpha\beta}$ are the selfdual generators of $\mathfrak{sl}(2,\C)$.
The isomorphism between pairs of spinor indices $(AA'), (BB'),\dots$ and internal Minkowski indices $\alpha,\beta,\dots$ is given by the identification of any four-vector $V^\alpha$ with the anti-Hermitian matrix
\begin{equation}
V^{AA'}=\frac{\I}{\sqrt{2}}\ou{\sigma}{AA'}{\alpha}V^\alpha,\quad V^\alpha=\frac{\I}{\sqrt{2}}\uo{\bar\sigma}{A'A}{\alpha}V^{AA'}.
\end{equation}

\section*{References}

\providecommand{\href}[2]{#2}\begingroup\raggedright\endgroup

\end{document}